\documentclass{article}
\usepackage{arxiv}

\usepackage[utf8]{inputenc} 
\usepackage[T1]{fontenc}    
\usepackage{hyperref}       
\usepackage{listings}
\usepackage{multirow}
\usepackage{xcolor}
\usepackage{amsmath}
\usepackage{makecell}
\usepackage{url}            
\usepackage{booktabs}       
\usepackage{amsfonts}       
\usepackage{nicefrac}       
\usepackage{microtype}      
\usepackage{amsmath,amsfonts}
\usepackage{algorithmic}
\usepackage[most]{tcolorbox}
\usepackage{booktabs}
\usepackage{xcolor}
\usepackage{tabularx}
\newcolumntype{C}[1]{>{\centering\arraybackslash}m{#1}}
\usepackage{amsmath, amssymb}
\usepackage{multirow}
\usepackage{adjustbox}
\usepackage{listings}
\usepackage{booktabs}
\usepackage{array}
\usepackage{makecell}
\usepackage{algorithm}
\usepackage{array}
\usepackage[caption=false,font=normalsize,labelfont=sf,textfont=sf]{subfig}
\usepackage{textcomp}
\usepackage{stfloats}
\usepackage{url}
\usepackage{verbatim}
\usepackage{graphicx}
\usepackage{cite}

\title{Enhancing LLM-Based Code Translation with Verified Multi-Semantic Representations}
\author{
 Yufu Wang \\
  School of Software\\
  Dalian University of Technology\\
  Dalian 116024, China \\
  \texttt{wyf@mail.dlut.edu.cn} \\
  \And
 He Jiang \\
  School of Software\\
  Dalian University of Technology\\
  Dalian 116024, China \\
  \texttt{jianghe@dlut.edu.cn} \
  \And
 Hao Lin \\
  School of Software\\
  Dalian University of Technology\\
  Dalian 116024, China \\
  \texttt{dlutlh@mail.dlut.edu.cn} \\
  \And
 Peiyu Zou \\
  School of Computer Science and Artificial Intelligence\\
  Liaoning Normal University\\
  Dalian 116029, China \\
  \texttt{zoupeiyu@lnnu.cn} \\
  \And
 Ang Jia \\
  School of Software\\
  Dalian University of Technology\\
  Dalian 116024, China \\
  \texttt{jiaang@dlut.edu.cn} \\
  \And
 Xiaochen Li \\
  School of Software\\
  Dalian University of Technology\\
  Dalian 116024, China \\
  \texttt{xiaochen.li@dlut.edu.cn} \\
  \And
 Zhilei Ren \\
  School of Software\\
  Dalian University of Technology\\
  Dalian 116024, China \\
  \texttt{zren@dlut.edu.cn} \\
}

\begin{document}
\maketitle
\begin{abstract}Large language models (LLMs) have shown great promise for automated code translation. However, existing approaches mainly rely on token-level statistical patterns for mapping, rather than generating target code based on a sufficient understanding of program semantics. As a result, translation outputs often contain intervenable logical and semantic errors. Such errors can potentially be mitigated by incorporating high-quality semantic guidance, such as functional descriptions and test cases. However, in real-world scenarios, these external semantic resources are often unavailable, making it necessary to construct semantic information directly from the source code to guide translation. This gives rise to two key challenges: first, how to construct semantic information that is sufficiently rich and diverse to characterize program behavior from multiple complementary perspectives; second, how to ensure that such semantics are accurate and reliable enough to effectively guide LLMs in code translation even in the absence of external resources. To address these challenges, we propose \textit{Multisage}, a multi-semantic augmentation and self-calibration framework for LLM-based code translation. \textit{Multisage} comprises three core modules. First, a semantic representation parsing module extracts structured base semantic representations from source code, including data-flow graphs, type-constraint analysis, and external API information. These representations capture intrinsic program semantics and provide the foundation for subsequent augmentation. Second, a multi-semantic augmentation module builds upon these base semantic representations to construct augmented semantics, such as function-level test cases, code summaries, and API-oriented descriptions and test cases, thereby expanding semantic coverage across different aspects of program behavior. Finally, a semantic consistency calibration module introduces semantics-preserving code mutations and cross-semantic consistency verification to automatically filter, calibrate, and refine the augmented semantics, ensuring their accuracy and reliability. Extensive experiments on the HumanEval-X code translation benchmark show that \textit{Multisage} improves translation success rates by up to 2.22× across diverse backbone models. It consistently outperforms vanilla prompting, instruction-tuned LLMs, and Chain-of-Thought (CoT) reasoning, with the largest relative gains observed on smaller models with limited capacity. These results demonstrate that explicit semantic augmentation effectively strengthens the reliability of LLM-based code translation.
\end{abstract}

\label{introduction}
Code translation refers to the automated process of converting source code from one programming language to another while preserving functional equivalence \cite{r1}, \cite{r2}. It plays a critical role in cross-language migration and software modernization, enabling developers to maintain and evolve large-scale systems more efficiently. As software ecosystems continue to expand and multiple programming paradigms coexist, code translation has become an essential component of software reengineering. In practice, many enterprise systems still rely on legacy programming languages whose limitations hinder maintainability, scalability, and security, leading to substantial migration and integration costs.

In this context, machine learning–based automatic code translation has emerged as a promising research direction \cite{avatar}, \cite{transcoder}, \cite{r3}. These approaches learn behavioral and functional constraints from source code and generate functionally equivalent constructs in the target language, enabling high-quality translation without extensive manual intervention \cite{r4}, \cite{correction}, \cite{VIM-PT}. Compared with rule-based systems or handcrafted conversion templates, learning-driven code translation substantially improves efficiency, reduces semantic deviation, and provides essential support for downstream tasks such as cross-language program understanding \cite{code_understanding}, automated bug repair \cite{code_repair}, and knowledge transfer \cite{knowledge_transfer}.

With the advent of the Transformer architecture \cite{transformer}, LLMs pre-trained on extensive code corpora have further advanced the field of automatic code translation. For example, CodeBERT \cite{codebert} achieves strong performance across multiple downstream code-related tasks through pre-training on large-scale code data. Meanwhile, TransCoder \cite{transcoder} combines unsupervised learning with back-translation and is trained on massive monolingual code datasets to acquire code translation capabilities. However, despite these advances, such approaches often struggle to capture key semantic constraints when handling complex cross-language differences, leading to logical or functional inconsistencies \cite{lost-in-translation}, \cite{correction}.

Recent studies have attempted to improve translation reliability by introducing additional semantic information to guide generation \cite{transcoder-st}, \cite{tree-tree}, \cite{unitrans}, \cite{alphatrans}. For example, Rozière et al. \cite{transcoder-st} incorporate language-agnostic semantic abstractions to enhance model understanding of source programs, while Yang et al. \cite{unitrans} employ automatically generated test cases to validate model outputs. Although these approaches provide stronger constraints for translation models, they typically address only localized failure modes and remain insufficient for handling complex semantic dependencies in real-world programs.

When program semantics involve intricate dependencies and significant cross-language differences, models still struggle to preserve logical consistency and functional equivalence. A closer analysis of translation outputs reveals that many failures are in fact intervenable, including logical inconsistencies, dependency misalignment, and data-parsing errors \cite{lost-in-translation}, \cite{correction}. These errors exhibit structured patterns rather than purely stochastic behavior, suggesting that they often stem from a common root cause: the lack of explicit and reliable semantic information during translation, as well as the absence of mechanisms to verify and calibrate such information. Consequently, models rely on token-level probabilistic associations for alignment, which can produce unstable outputs in complex scenarios \cite{lost-in-translation}, \cite{token_level}. To mitigate this issue, prior work introduces external semantic resources such as test cases, documentation, or formal specifications to strengthen translation constraints \cite{unitrans}, \cite{transcoder-st}. However, such resources are often unavailable in real-world software systems. Therefore, improving the reliability of LLM-based code translation requires systematically constructing and enforcing key semantic constraints directly from source code. This requirement gives rise to two key challenges:

\begin{itemize}
    \item{\textbf{Constructing multi-semantic information}: In program analysis and code understanding, semantics refers to information that characterizes program behavior, logical relations, and functional constraints \cite{sem-1}. Such information directly determines the functional correctness and consistency of translation results \cite{sem-2, sem-5}. Prior studies \cite{sem-3, sem-4} show that different semantic perspectives capture complementary aspects of program behavior, including control relations along execution paths, value propagation, and constraints governing function and API usage. However, these perspectives differ in representation and focus, and any single semantic form typically captures only a limited facet of program behavior. Consequently, relying on a single semantic representation makes it difficult to cover the diverse semantic constraints required across execution contexts. A key challenge is therefore how to systematically construct multiple complementary semantic forms directly from source code to improve overall semantic coverage.}

    \item{\textbf{Ensuring the accuracy and reliability of semantic information}: Semantic information is often constructed through static inference, approximate assumptions, or limited contextual evidence, making it susceptible to noise, incompleteness, and bias \cite{sem-2}. Different semantic perspectives may therefore introduce conflicting or inconsistent constraints. When such imperfect semantics are directly used to guide translation models, they may fail to correct errors and even mislead the model to produce outputs that appear plausible but deviate from the intended behavior \cite{tit, xlcost}. A key challenge is thus how to systematically verify, filter, and calibrate semantic information so that it provides reliable guidance during translation.}
\end{itemize}

To address these challenges, we propose \textit{Multisage}, a multi-semantic augmentation and self-calibration framework for LLM-based code translation. \textit{Multisage} consists of three core modules. First, a semantic representation parsing module extracts structured semantic units from source code through static analysis. Second, a multi-semantic augmentation module generates complementary semantic views, including code summaries, API usage information, and function-level test cases, to enhance semantic coverage. Third, a semantic consistency calibration module performs semantics-preserving mutations and cross-view consistency verification to filter unreliable semantic signals. Together, these modules construct, verify, and calibrate semantic information, enabling LLMs to utilize not only richer semantics but also explicitly modeled and reliable semantic guidance during translation.

We conduct a systematic study on the HumanEval-X benchmark using LLMs with diverse architectures and parameter scales to evaluate the effectiveness and cross-model generalization of \textit{Multisage}. Under vanilla prompting, \textit{Multisage} improves translation success rates by up to 2.22×, with consistent gains across all evaluated models. We further compare \textit{Multisage} with representative specialized code translation models, such as TransCoder, under identical evaluation settings. \textit{Multisage} achieves substantially higher CodeBLEU scores while maintaining competitive execution success rates, indicating that explicit multi-semantic augmentation provides more comprehensive implementation-level semantic constraints than approaches relying primarily on structural modeling or model-specific design. Compared with alternative semantic enhancement strategies such as CoT reasoning and single-stage semantic prompting, \textit{Multisage} consistently improves translation success rates by up to 1.42× on small-scale models, 1.28× on mid-scale models, and 1.17× on large-scale models. Overall, these results demonstrate that \textit{Multisage} provides strong and stable performance gains across diverse model configurations, effectively reducing intervenable semantic errors and enhancing the reliability of LLM-based code translation.

The main contributions of this work are summarized as follows:

\begin{itemize}
    \item{We identify and analyze a key limitation of existing LLM-based code translation: most failures are intervenable semantic errors caused by the lack of explicitly modeled and reliable semantic constraints, rather than insufficient model capacity or randomness. Detailed error categorization and quantitative statistics are presented in Section~\ref{motivation}.}
    \item{We propose \textit{Multisage}, a multi-semantic augmentation and self-calibration framework for LLM-based code translation. \textit{Multisage} integrates three cooperative components: semantic representation parsing, multi-semantic augmentation, and semantic equivalence tuning, to systematically construct diverse and complementary semantic forms solely from source code. The framework constructs and calibrates diverse semantic representations from source code to provide reliable and controllable semantic guidance for translation.}
    \item{We conduct comprehensive evaluations on HumanEval-X with multiple LLMs of varying scales, showing that \textit{Multisage} consistently improves translation success rates across models, achieving gains of up to 2.22×, and outperforming representative semantic enhancement strategies.}
\end{itemize}

\section{Background and Motivation}
\label{Background and Motivation}
\subsection{LLM-Based Code Translation}
Automatic code translation aims to convert programs from one programming language into semantically equivalent implementations in another \cite{avatar}, \cite{tree-tree}. With the rapid development of LLMs, recent LLM-based approaches have significantly advanced the state of the art in code translation \cite{codebert}, \cite{transcoder}. By leveraging large-scale pretraining on code corpora and strong generative capabilities, LLMs can produce target-language code that is syntactically fluent and stylistically consistent, often resembling human-written implementations \cite{FSCTrans}, \cite{VIM-PT}.

Despite their impressive performance, LLM-based code translation systems face fundamental challenges inherent in the generative nature of LLMs. Code translation is a strictly semantics-preserving task, where even subtle deviations in control flow, data dependencies, or API usage may result in incorrect program behavior. However, LLMs primarily model token-level probability distributions and lack built-in mechanisms for explicit semantic verification. As a result, they may generate translations that appear plausible on the surface but violate essential semantic constraints \cite{xlcost}.

Most existing LLM-based code translation methods are built upon large-scale code corpora, from which they learn statistical correspondences between source and target languages. While this data-driven paradigm works well for common translation patterns, it often struggles when accurate translation requires a deeper understanding of program intent, execution behavior, or the semantics of external libraries. As a result, LLMs may generate translations that are syntactically valid but semantically incorrect, especially for programs involving complex logic, conditional structures, or non-trivial API interactions \cite{lost-in-translation}. Importantly, such errors arise not only from model capacity limitations but also from the absence of explicit mechanisms for exposing and enforcing semantic constraints during generation \cite{correction}, \cite{unitrans}.

\subsection{Semantic Information}
Program semantics describes the meaning and behavior of a program, including how inputs are processed, how internal states evolve, and what outputs are produced \cite{sem-1}, \cite{sem-2}. Prior studies show that program semantics can be represented through multiple complementary forms that capture different aspects of program behavior \cite{sem-2}.

Among these forms, natural-language code summaries provide high-level descriptions of program functionality and developer intent, which have been shown to improve downstream tasks such as code search and understanding \cite{sem-2}, \cite{sem-3}. In addition, test cases serve as practical carriers of semantic constraints by explicitly specifying input–output behaviors and expected program functionality \cite{sem-4}, \cite{sem-5}.

For LLM-based code generation and translation, such semantic information helps bridge the gap between surface-level code patterns and deeper program understanding. By providing explicit semantic signals such as summaries, API properties, and test cases, models can better align generated code with the intended functionality. This capability is particularly important for code translation, where semantic preservation is required.

\subsection{Motivation}
\label{motivation}

\begin{figure*}[!t]
\centering
\subfloat[]{\includegraphics[width=2in]{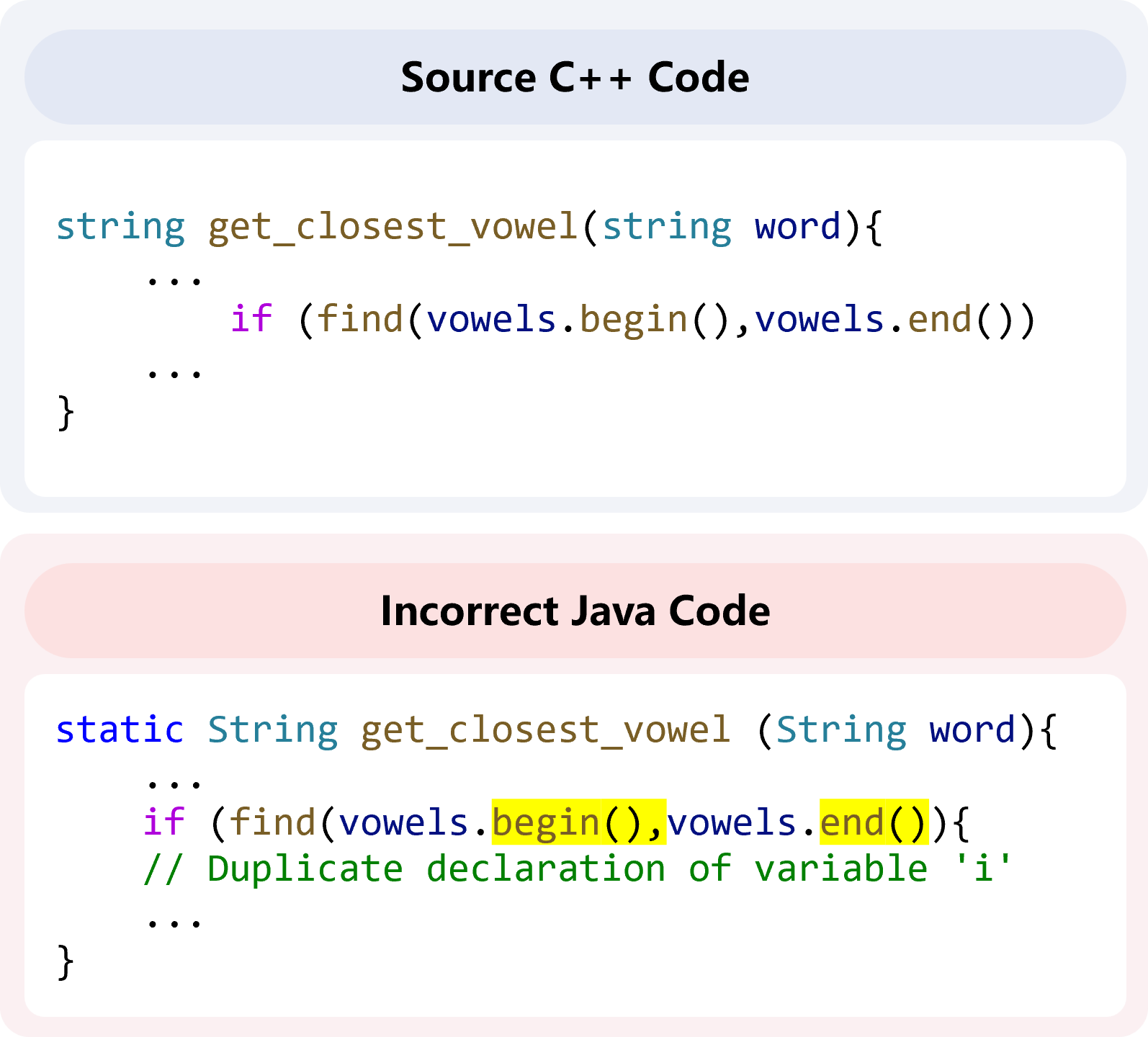}%
\label{fig_2_1_first_case}}
\hfil
\subfloat[]{\includegraphics[width=2in]{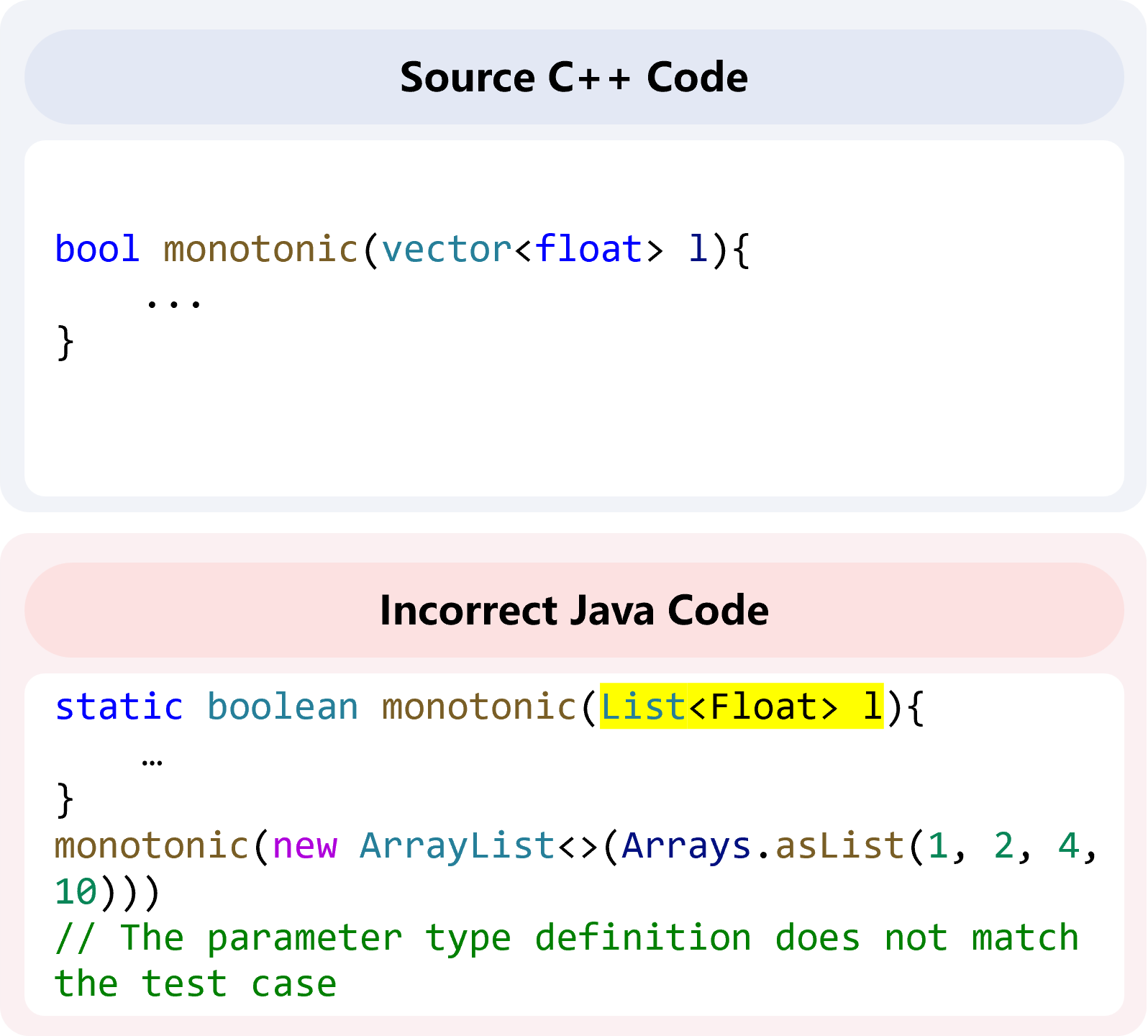}%
\label{fig_2_1_second_case}}
\hfil
\subfloat[]{\includegraphics[width=2in]{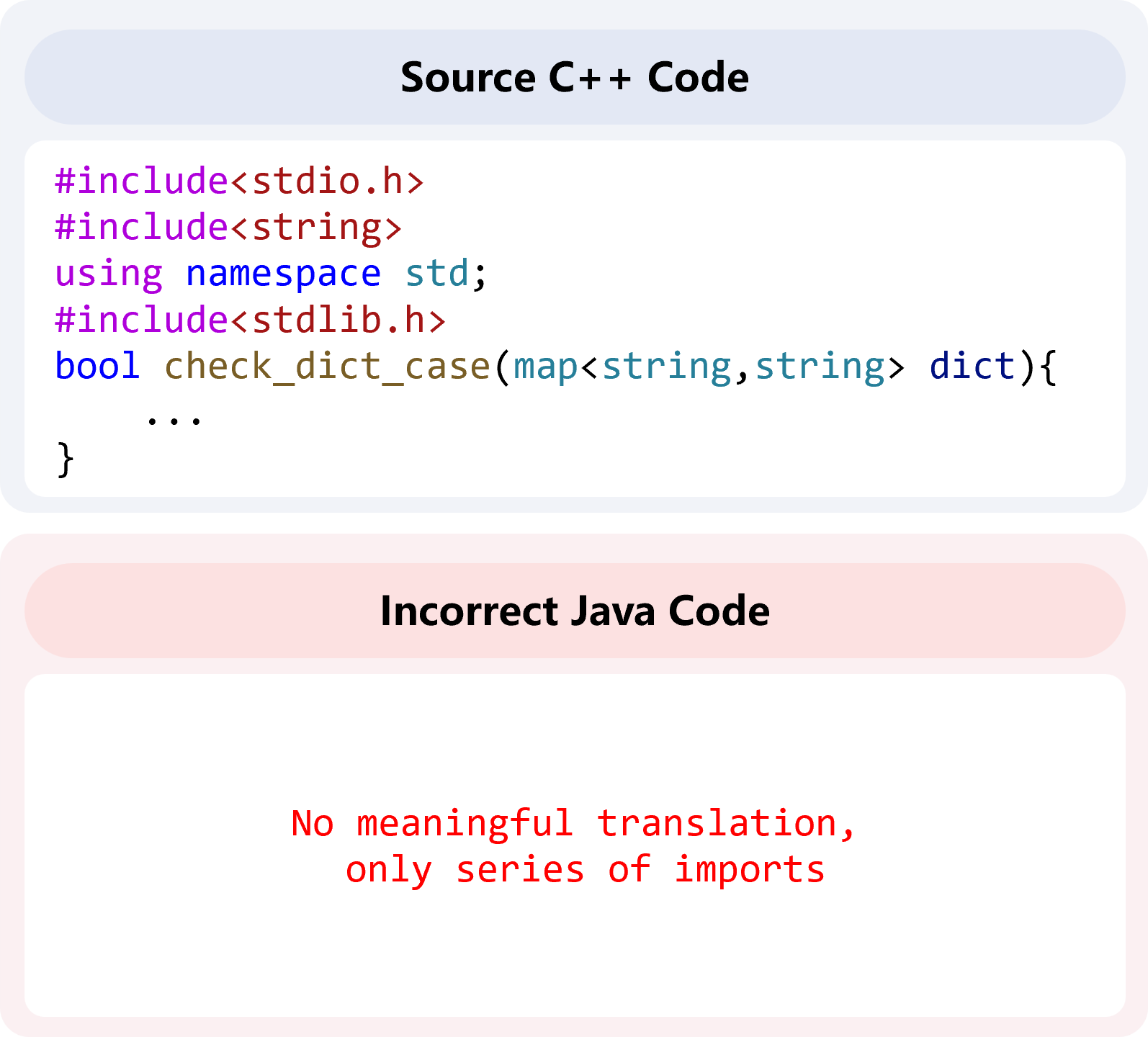}%
\label{fig_2_1_third_case}}
\caption{Categories of LLM-Based Code Translation Failures. (a) Dependency and Logical Errors. (b) Data Parsing Errors. (c) Model-Specific Errors.}
\label{moti_1}
\end{figure*}
To better understand the limitations of existing LLM-based code translation approaches, we conduct a systematic empirical analysis of translation outputs across multiple models. The results indicate that a substantial proportion of failures can be attributed to intervenable semantic errors, which typically arise from missing or misinterpreted semantic constraints. Meanwhile, a smaller portion of failures is associated with model capacity limitations or stochastic decoding behaviors. These findings suggest that improving the reliability of code translation hinges on systematically introducing explicit and verifiable semantic constraints.

To further analyze the structure of these failures, we categorize common failure modes into three types, as illustrated in Fig. \ref{moti_1} \cite{correction, lost-in-translation}. The first category, dependency and logical errors, includes issues such as unresolved dependencies, incorrect API usage, or flawed control-flow reasoning. The second category, data parsing errors, involves structural or type-level inconsistencies, such as mismatched return formats or invalid type handling. These two categories correspond to intervenable translation errors, as they primarily arise from missing or misinterpreted semantic constraints and can, in principle, be mitigated through appropriate semantic guidance or auxiliary analysis \cite{correction}, \cite{lost-in-translation}. In contrast, the third category, model-specific errors, refers to meaningless or ill-formed outputs produced by the model itself, which cannot be effectively corrected through semantic constraints and are largely determined by the model architecture and decoding process.

\begin{figure}[!t]
\centering
\includegraphics[width=3.5in]{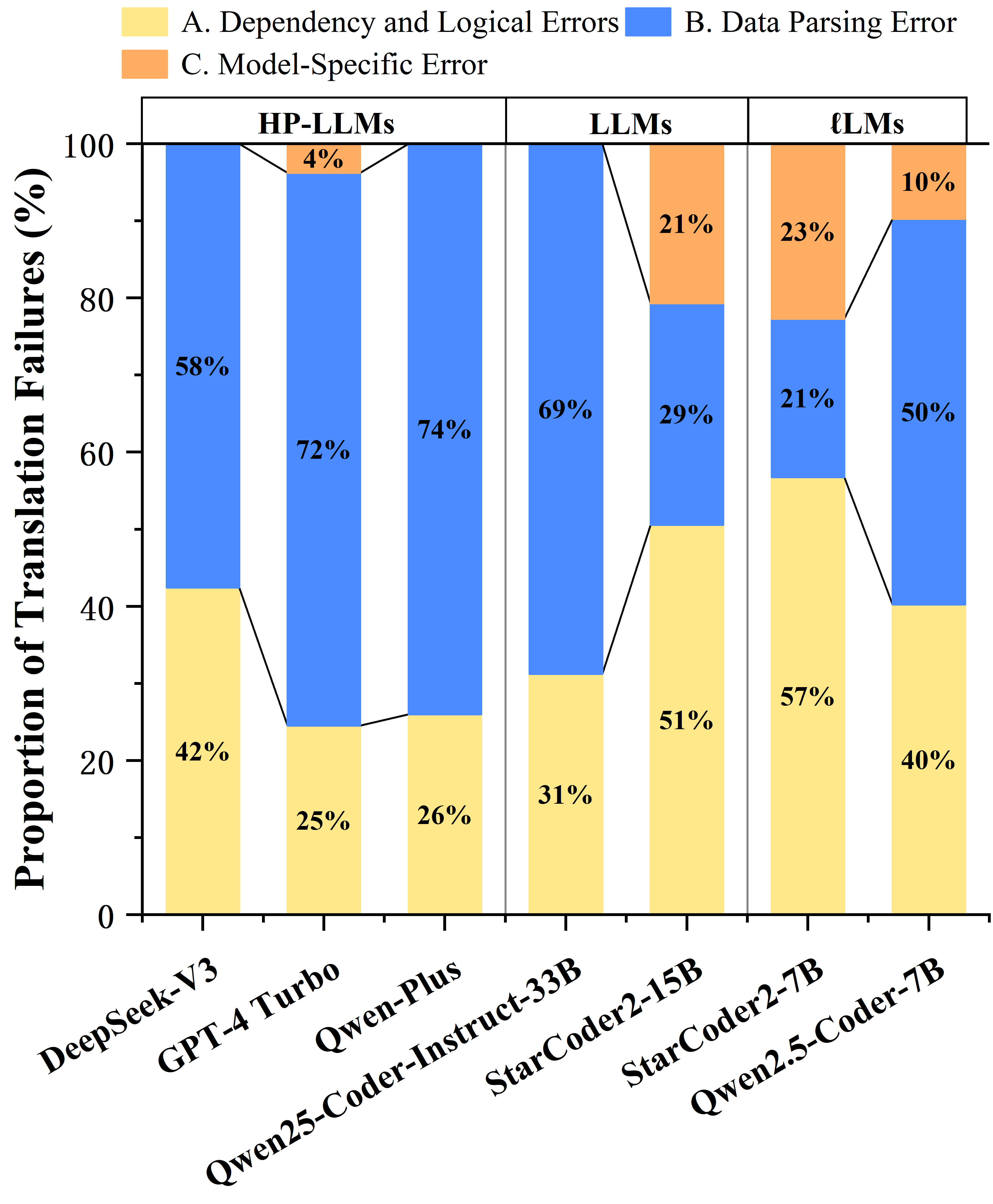}
\caption{Proportion of Code Translation Failure Types Across LLM Scales. Percentages are rounded to the nearest integer and may not sum to exactly 100\%.}
\label{moti_2}
\end{figure}

To quantify the prevalence of these error types, we analyze models with different parameter scales and architectures. As shown in Fig. \ref{moti_2}, models are grouped into three categories: high-performance large models (HP-LLMs, $>$100B parameters), mid-scale models (MS-LLMs, 10B–100B), and lightweight models ($\ell$LMs, $<$10B). Across HP-LLMs and most large models, the majority of translation failures fall into the intervenable category, whereas model-specific errors account for only a small fraction. Even for lightweight models, model-specific errors typically represent only 10\%–23\% of failures, indicating that most translation errors are potentially correctable.

\begin{figure}[!t]
\centering
\subfloat[]{\includegraphics[width=3in]{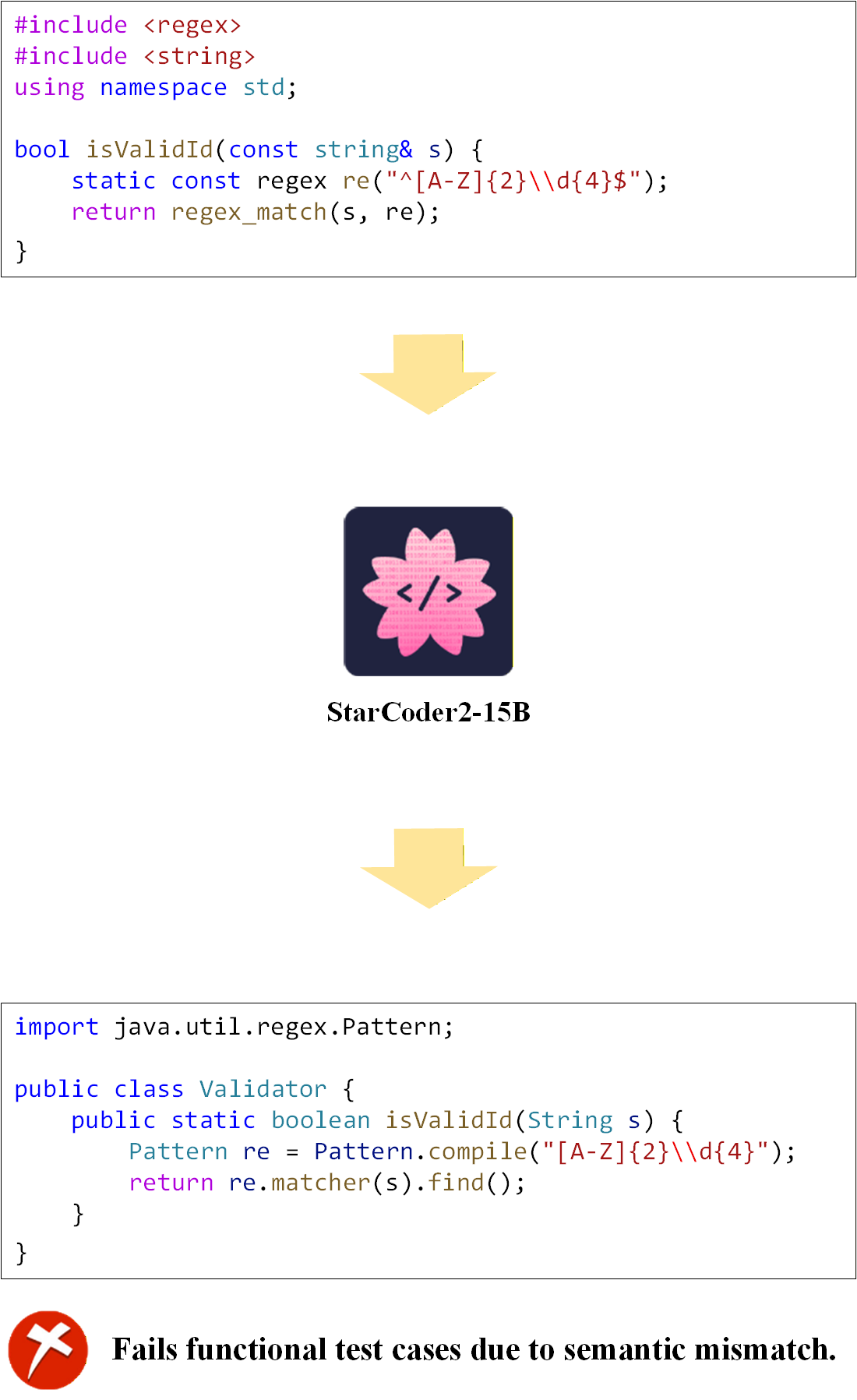}%
\label{fig_2_3_first_case}}
\hfil
\subfloat[]{\includegraphics[width=3in]{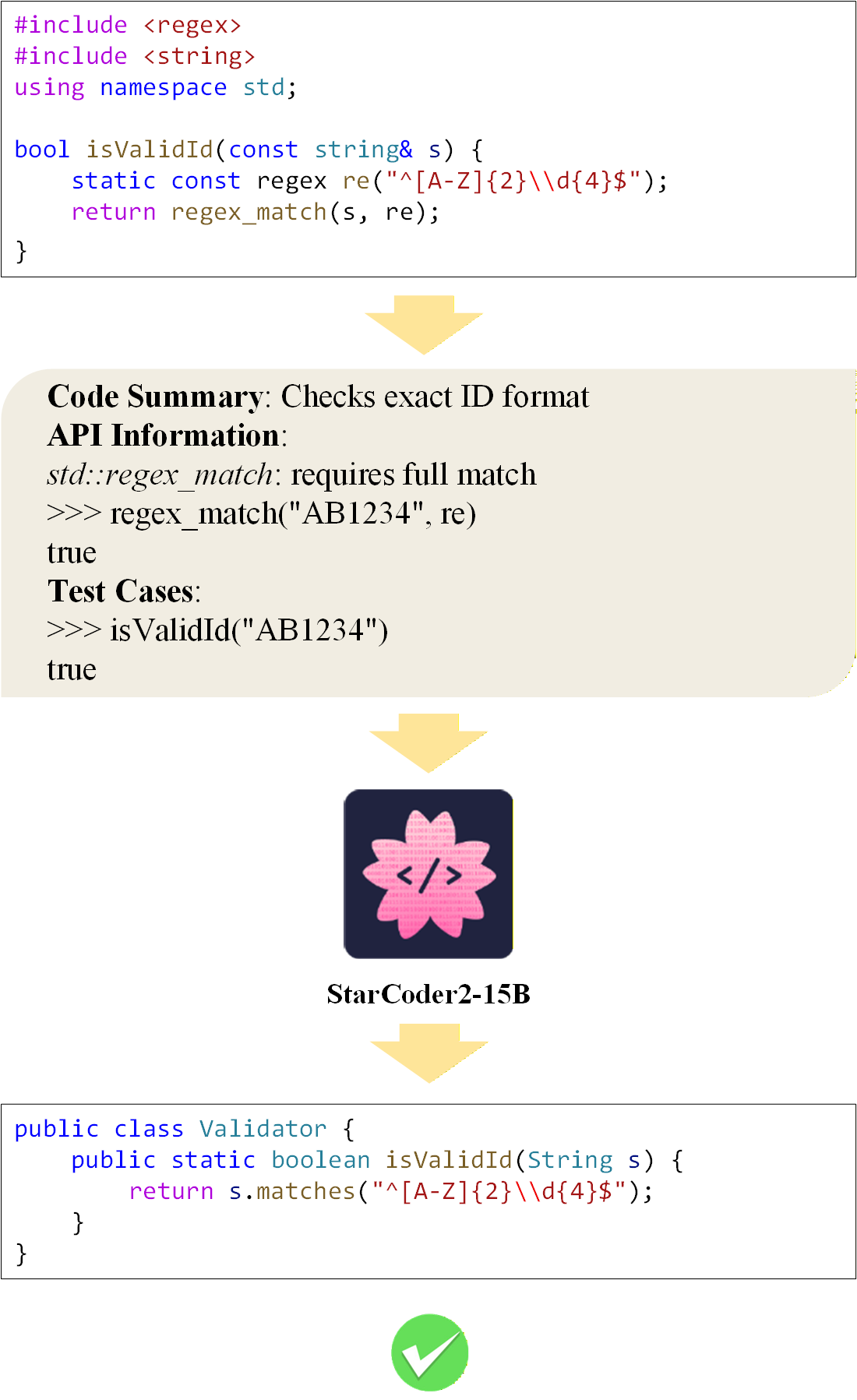}%
\label{fig_2_3_second_case}}
\caption{Impact of Explicit Semantic Information on Code Translation. (a) Translation without Explicit Semantic Guidance. (b) Translation with Explicit Semantic Guidance.}
\label{moti_3}
\end{figure}

Fig. \ref{moti_3} provides a concrete example. When translating a C++ function that validates identifiers using \texttt{std::regex\_match}, a vanilla StarCoder2-15B model generates syntactically valid Java code but performs partial pattern matching instead of enforcing a full match. By providing explicit semantic guidance such as a code summary, API semantics specifying full-match requirements, and minimal test cases, the model produces a semantically correct translation. This example illustrates that many failures arise not from code generation ability but from the absence of explicit semantic constraints.

Taken together, these observations point to a key insight: improving LLM-based code translation therefore does not primarily require larger models, but rather mechanisms that expose and leverage semantic information during translation. Since many failures are intervenable, incorporating explicit semantic signals offers a practical path toward more reliable and semantics-preserving code translation.

While these findings highlight the importance of semantic grounding, simply exposing semantic information is insufficient. Program semantics are inherently multi-faceted, encompassing functional intent, API constraints, type relations, and control dependencies. A single representation therefore, cannot fully capture the complex behavior of real-world programs. A key challenge is how to systematically derive complementary semantic representations from source code to provide comprehensive guidance for translation.

Another challenge concerns the reliability of constructed semantics. Because semantic information is often derived through approximate inference, it may contain noise or inconsistencies. Without proper filtering and calibration, such imperfect semantics may mislead the model. Therefore, effective semantic mediation requires mechanisms to both construct diverse semantic representations and verify their reliability. In the next section, we present such a framework, which operationalizes multi-semantic construction and self-calibration to provide reliable semantic guidance for code translation.

\begin{figure*}[h!]
\centering
\includegraphics[width=6in]{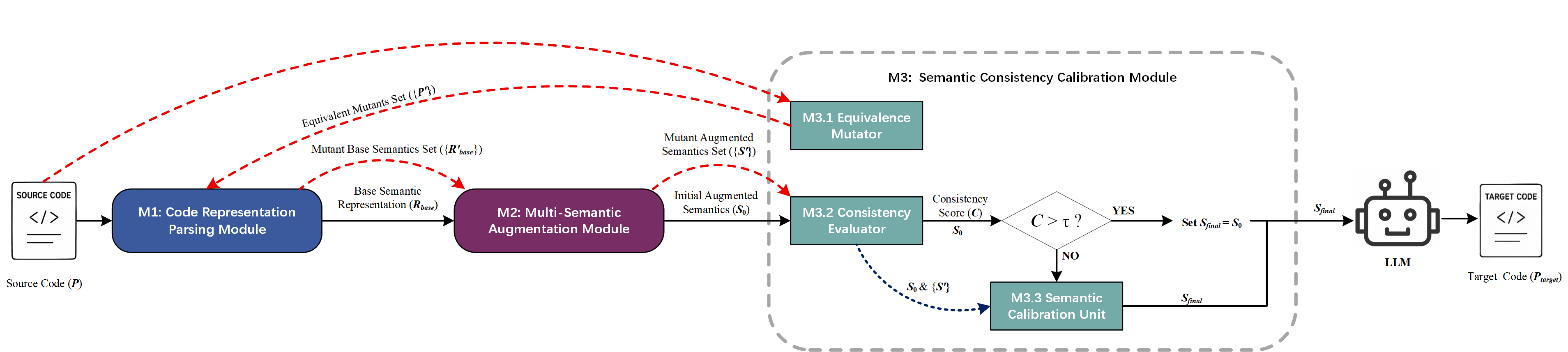}
\caption{Workflow of Multisage.}
\label{overview}
\end{figure*}
\section{Framework}
\label{Framework}
\subsection{Overview}
\textit{Multisage} is a multi-semantic augmentation and self-calibration framework for LLM-based code translation. As illustrated in Fig.~\ref{overview}, it consists of three modules: semantic representation parsing, multi-semantic augmentation, and semantic consistency calibration. Given a source-language function, the framework first reconstructs structured semantic representations from the code, then generates diverse semantic views, and finally calibrates these semantics before providing them together with the original code to the LLM as semantically grounded guidance for translation.

The semantic representation parsing module performs static analysis on the input program to construct a language-agnostic semantic representation that captures control flow, type constraints, and API dependencies. This structured base semantic representation serves as the foundation for subsequent semantic augmentation and calibration.

The multi-semantic augmentation module then derives complementary semantic views from the parsed base semantic representations, including code summaries, API semantic descriptions, and function-level test cases. These heterogeneous views capture program behavior, constraints, and usage intent from different perspectives, improving the model’s ability to preserve semantic dependencies during translation.

Finally, the semantic consistency calibration module improves the reliability of the constructed semantics through semantics-preserving code mutations and cross-view consistency checking. It identifies and filters unreliable semantic signals and organizes the calibrated semantics into a structured prompt together with the original code. This process provides stable semantic guidance for the LLM and improves the functional correctness and robustness of translation.

\subsection{M1: Code Representation Parsing Module}

The code representation parsing module aims to structurally parse and reconstruct semantic representations from source code, focusing on three explicit and machine-interpretable semantic aspects: control flow, type constraints, and external interfaces. The module performs static analysis to generate structural templates and dependency descriptions, which provide logical foundations and structural constraints for subsequent semantic augmentation and cross-language code generation. The resulting set of structured semantics is collectively referred to as the basic semantic representation (\(R{_{base}}\)), which serves as the input to the subsequent multi-semantic augmentation module.

\textbf{Control Flow Extraction}. Control flow forms the backbone of a function’s logic, capturing the execution order and branching behavior among statements. To extract this structure, we perform static analysis over the source program using an abstract syntax tree (AST) parser\footnote{https://tree-sitter.github.io/tree-sitter} and a control flow graph (CFG) generator\footnote{https://github.com/joernio/joern}, where the AST provides structural boundaries for statements and control constructs, and the CFG captures the corresponding execution paths. The CFG captures the control-flow structure of program execution, where nodes correspond to statement blocks and edges denote execution paths, including conditional branches, loops, and exception handling. The extracted CFG is then linearized into a structured execution template using block-ordering techniques, which is included as one component of the \(R{_{base}}\). The linearization follows a depth-first traversal over the CFG, where conditional and loop structures are explicitly encoded as structured control tokens.

For example, consider the following C++ function:
\begin{lstlisting}[label={model1-example1}, basicstyle=\ttfamily\footnotesize, frame=single]
int maxValue(int a, int b) {
    if (a > b) return a;
    else return b;
}
\end{lstlisting}
Its control flow can be represented as:
\begin{lstlisting}[basicstyle=\ttfamily\footnotesize, frame=single, escapeinside={(*@}{@*)}]
Entry (*@$\rightarrow$@*) Condition(a > b) (*@$\rightarrow$@*) {True: return a | False: 
return b} (*@$\rightarrow$@*) Exit
\end{lstlisting}

\textbf{Type Constraint Parsing}. Since our focus is on preserving functional semantics rather than implementation details, we extract only the function interface and its input–output types. Using the AST derived from the parser, we analyze function signatures to identify formal parameters and return types. These constraints are encoded into a structured type-constraint representation and injected into the LLM prompt to regulate variable declarations and return-value inference during code generation, thereby encouraging type-consistent generation and improving semantic alignment. 

\textbf{External Interface Extraction}. External interfaces define how a function interacts with its surrounding environment, including library dependencies, API invocations, and input–output behaviors that are not fully captured by internal control or type structures. To model these aspects, we analyze function bodies using the AST to identify external symbols, such as library function calls, object method invocations, and imported modules. 

These elements are further normalized into a structured interface representation, where each external interaction is abstracted into a canonical form consisting of the invoked entity, its arguments, and its functional role (e.g., I/O operation, container manipulation, or numerical computation). This abstraction reduces language-specific variability while preserving essential semantic intent.

The extracted interface information is incorporated into \(R_{\text{base}}\) as a complementary semantic component, enabling the model to better align external behaviors across programming languages. By explicitly exposing these dependencies, the model is guided to generate target code that preserves critical interactions with external systems and libraries.

\subsection{M2: Multi-Semantic Augmentation Module}

The multi-semantic augmentation module aims to enrich the base semantic representation \(R_{\text{base}}\) by introducing diverse and complementary semantic signals that are not fully captured by structural analysis alone. Given \(R_{\text{base}}\) as input, this module constructs and refines multiple forms of explicit semantic information, including functional summaries, test cases, and API-level descriptions, to provide comprehensive semantic guidance for code translation. 

\begin{figure}[!t]
\centering
\includegraphics[width=6in]{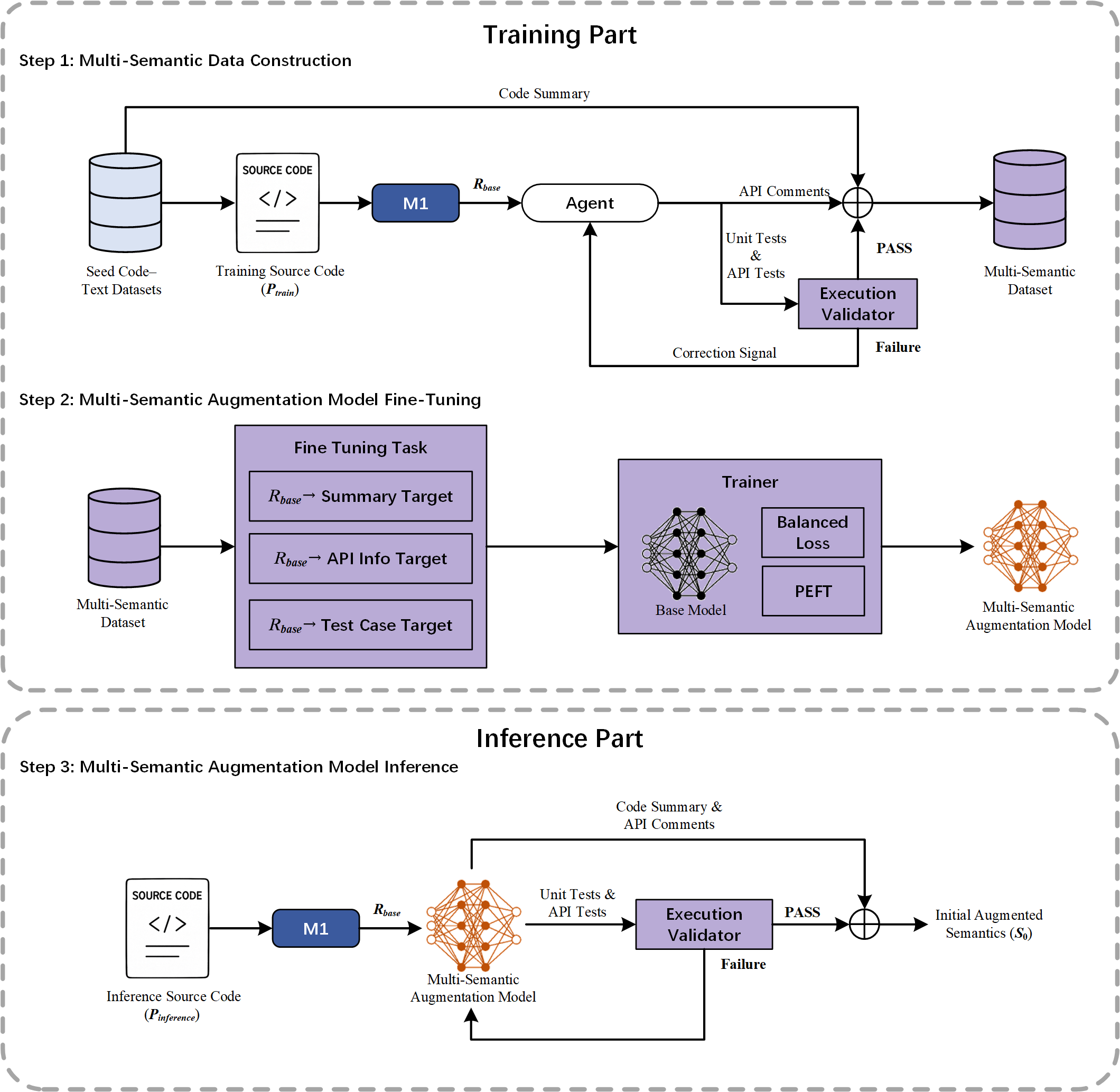}
\caption{Workflow of Multi-Semantic Augmentation Module.}
\label{module2_overview}
\end{figure}

The overall workflow of this module is illustrated in Fig.~\ref{module2_overview}. Specifically, the module operates in three stages: (1) multi-semantic data construction, where diverse semantic views are generated and validated; (2) multi-semantic augmentation model fine-tuning, where a unified model is trained to capture cross-semantic relationships; and (3) multi-semantic augmentation model inference, where multiple semantic signals are jointly produced and iteratively refined. The resulting augmented semantics serve as a richer and more reliable semantic representation for subsequent semantic consistency calibration.

\textbf{Multi-Semantic Data Construction}. Due to the lack of high-quality datasets that comprehensively cover multiple semantic dimensions of source code, existing large language models often struggle to learn complex code semantics sufficiently from a single supervision signal. To address this limitation, we propose an LLM-based automatic multi-semantic data construction approach, whose overall workflow is illustrated as Step 1 in Fig. \ref{module2_overview}. We first adopt the composite code datasets XLCoST \cite{xlcost} and XCodeEval \cite{xcodeeval} as the raw seed code–text datasets, both of which provide natural-language functional requirements paired with corresponding code snippets and thus serve as structured input–output templates for multi-task modeling. We then treat the source code \(P\) as the core modeling unit and, based on the base semantic representation \({R_{base}}\) produced by the code representation parsing module, leverage an LLM-based semantic generation agent to construct a Multi-Semantic Dataset that covers diverse semantic perspectives, including function-level code summaries, function-level test cases, and API-level semantic descriptions with associated test cases.

To ensure the reliability and task consistency of the generated semantic information, we further introduce an execution validator that performs executability checking over the generated function-level and API-level test cases. Samples that pass validation are directly added to the Multi-Semantic Dataset, whereas those that fail trigger the generation of a Corrective Signal, which drives multi-round semantic refinement and correction. Guided by this feedback signal, the LLM regenerates or locally adjusts the corresponding semantic outputs, enabling test-case–like samples to gradually converge toward semantically consistent and logically complete states across different semantic perspectives. Representative prompting templates for semantic generation and refinement are provided in Appendix~\ref{appendixA}.

\textbf{Multi-Semantic Augmentation Model Fine-Tuning}. To achieve cross-semantic sharing and dynamic optimization balancing, \textit{Multisag}e introduces a multi-task joint fine-tuning mechanism inspired by FAMO \cite{famo} and MFTCoder \cite{MFTCoder} during the training of the multi-semantic augmentation model. This mechanism adaptively adjusts the loss weights of different semantic tasks throughout training, thereby dynamically coordinating their convergence speeds and improving the overall stability and robustness of optimization. Unlike traditional single-task fine-tuning or naive multi-task training, \textit{Multisage} treats each semantic task as an independent optimization objective, which is jointly optimized within a shared parameter space. 

Assume there are \(N\) semantic generation tasks, and each task \({T_i}\) is associated with a training dataset \({D_i} = \{ (x_i^{(j)},y_i^{(j)})\} _{j = 1}^{{M_i}}\), where \(x_i^{(j)}\) denotes the \(j\)-th input sample drawn from the Multi-Semantic Dataset, \(y_i^{(j)}\) represents the corresponding task-specific target, and \({{M_i}}\) is the total number of samples for task \({T_i}\). The overall optimization objective of \textit{Multisage} is defined as a weighted multi-task loss:

\begin{equation}
\label{eq1}
{\cal L}(\theta ) = \sum\limits_{i = 1}^N {{w_i} \cdot } {{\cal L}_i}(\theta )
\end{equation}

where \({{\cal L}_i}(\theta )\) is the loss function of the \(i\)-th task with model parameters denoted as \(\theta \), and \({w_i}\) denotes the \(i\)-th normalized adaptive task weight, which dynamically balances the importance and convergence rate of different tasks during training. The adaptive weights \(w_i\) are updated according to the relative improvement of each task during training, enabling automatic coordination of task convergence.

Detailed formulations of the weighting strategy and optimization procedure are provided in Appendix~\ref{appendixB}.

\textbf{Multi-Semantic Augmentation Model Inference}. During inference, given a source code \(P\), we first obtain its base semantic representation \({R_{base}}\) from the code representation parsing module and feed \({R_{base}}\) into the multi-semantic augmentation model. The model then generates multiple forms of explicit semantic outputs in parallel, including function-level code summaries, API-level semantic descriptions, API-level test cases, and function-level test cases.

To ensure the practical usability of the generated semantics, the API-level and function-level test cases are automatically submitted to an execution validator, which only checks whether the test cases can be successfully compiled and executed. If a test case passes validation, the corresponding semantic output is retained; otherwise, the system triggers regeneration of the test cases for the associated function or API. During this multi-round refinement process, only test cases related to failed functions or APIs are regenerated, while previously validated outputs remain unchanged. The generate–validate loop is executed for at most three refinement rounds. If validation still fails after three rounds, further refinement is skipped, and the sample is marked as a low-confidence instance to be handled uniformly by the subsequent semantic consistency calibration module.

\subsection{M3: Semantic Consistency Calibration Module}

To further improve the accuracy and reliability of the constructed semantic guidance, \textit{Multisage} introduces a semantic consistency calibration module. The core idea is that semantically equivalent programs should induce highly consistent semantic representations, even if their surface forms differ. Therefore, instead of directly trusting the initial augmented semantics \({S_{0}}\), we evaluate its stability by comparing it against semantics reconstructed from semantics-preserving program variants.

Concretely, given the source code \(P\), the equivalence mutator generates an equivalent mutant set  \(\{ P'\} \). Each mutant \(P'\) is then processed by the modules M1 and M2, to obtain its mutant base semantics set \(R{'_{base}}\) and the corresponding mutant augmented semantics set \(\{ S'\} \). The consistency evaluator computes a consistency score \(C\) between the \({S_{0}}\) and \(\{ S'\} \). When \(C\) exceeds a predefined threshold \(\tau \), we regard \({S_{0}}\) as sufficiently reliable and directly use it as the final semantic guidance. Otherwise, the semantic calibration unit is activated to selectively integrate the most consistent semantic components across \({S_{0}}\) and \(\{ S'\} \), producing the calibrated final semantics \({S_{final}}\). This mechanism enables \textit{Multisage} to automatically filter noisy or contradictory semantics and provide the LLM with stable, semantically grounded guidance for code translation.

\textbf{Equivalence Mutator}. The equivalence mutator generates a small set of semantics-preserving program variants to assess the stability of augmented semantics. 

The underlying principle is that semantically equivalent programs, despite syntactic differences, should induce consistent semantic representations. 
Therefore, instability across equivalent variants indicates potential noise or overfitting in the constructed semantics.

Given a source program $P$, the mutator produces a set of equivalent variants $\{P'\}$ through static rule-based transformations applied to its AST and CFG representations. 
These transformations are designed to preserve program behavior for all valid inputs.

We instantiate the mutator along three transformation dimensions: 1) \textit{Expression-level rewrites}, which apply algebraic or logically equivalent transformations to side-effect-free expressions; 2) \textit{Control-flow–preserving rewrites}, which restructure conditional and loop constructs while maintaining equivalent execution paths; 3) \textit{API-level substitutions}, which replace library calls with verified equivalent alternatives under identical preconditions.

All transformations are purely static and deterministic, operating on AST, CFG, and symbol-table representations without learned components. 
Variants that fail compilation or execution checks are discarded by the downstream validator, ensuring that only behaviorally consistent mutants are retained. The complete set of transformation rules is provided in Appendix~\ref{appendixC}. In practice, we generate 3–5 equivalent variants for each input program.

\textbf{Consistency Evaluator}. The consistency evaluator measures the stability of the initial augmented semantics \({S_{0}}\) by comparing it against the mutant augmented semantics set \(\{ S'\} \) reconstructed from the equivalent mutants. Our design is grounded in a well-established conclusion in \cite{eq_mut_1} that equivalent mutants behave identically to the original program for all possible inputs, and thus preserve the same observable semantics despite syntactic differences. Prior work \cite{eq_mut_3} on program semantics and code representation learning further argues that semantic representations should remain invariant under semantics-preserving program transformations, rather than being tied to surface-level syntax. Motivated by these findings, we assume that semantically equivalent programs should induce highly consistent semantic representations, and large deviations mainly indicate instability or noise in the constructed semantics.

Directly requiring natural-language descriptions in \({S_{0}}\) and \(\{ S'\} \) to be exactly identical is, however, unrealistic, since semantically equivalent descriptions may differ in wording or structure. To address this, we normalize each semantic output into a set of semantic units \(U(S)\), where a unit corresponds to an API-usage constraint, a functional intent statement, or a behavioral assertion in a test case. We then compare semantic units based on token-level similarity rather than exact string matching.

For each semantic unit \(u \in U({S_0})\), we define the set of mutants in which \(u\) can be matched as:
\begin{equation}
    \label{eq8}
    {
\mathcal{M}(u)
=
\left\{
S' \in \{S'\}
\;|\;
\exists v \in U(S')
\text{ s.t. }
\mathrm{sim}(u,v)\ge\delta
\right\}
    }
\end{equation}

The similarity function \(sim(u,v)\) is implemented as token-level Jaccard similarity:

\begin{equation}
    \label{eq9}
\mathrm{sim}(u,v)
=
\frac{|T(u)\cap T(v)|}{|T(u)\cup T(v)|}
\end{equation}

with \(T(u)\) and \(T(v)\) denoting the sets of normalized tokens extracted from \(u\) and \(v\), respectively. Token normalization includes lowercasing, stop-word removal, and preserving identifiers, API names, and key action verbs. This lightweight metric has been widely used in software-engineering text analysis, and allows us to measure semantic overlap without introducing additional neural models. We introduce a similarity threshold \(\delta \) to determine whether two semantic units are considered matched, i.e., they are treated as aligned only when their informative tokens sufficiently overlap.

The support of semantic unit \(u\) is defined as the fraction of mutants in which \(u\) can be matched:

\begin{equation}
    \label{eq10}
\mathrm{supp}(u)
=
\frac{|\mathcal{M}(u)|}{|\{S'\}|}
\end{equation}

We then define the stable semantic unit set as
\begin{equation}
    \label{eq11}
U^{\mathrm{stable}}
=
\left\{
u \in U(S_{0})
\;\middle|\;
\mathrm{supp}(u)\ge\rho
\right\}
\end{equation}

where \(\rho\) is the majority-support threshold. We adopt a majority-support threshold of \(\rho = 0.5\) when determining whether a semantic unit is stable. This design follows the classical majority-vote principle: if more than half of the equivalent mutants reproduce the same semantic unit (under similarity threshold \(\delta \)), the unit is highly likely to reflect the true program semantics rather than noise introduced during augmentation.

Finally, the consistency score \(C \in [0,1]\) between the \({S_{0}}\) and \(\{ S'\} \) is computed as:

\begin{equation}
    \label{eq12}
C=\frac{|U^{\mathrm{stable}}|}{|U(S_{0})|}
\end{equation}

which quantifies the proportion of semantic units in \({S_{0}}\) that are consistently supported by the equivalent mutants under similarity threshold \(\delta \).

In all experiments, we set the consistency threshold \(\tau \) to 0.6. This value reflects the following rationale: each semantic unit must be confirmed by the majority of equivalent mutants to be considered stable, while \(\tau = 0.6 \) further requires that most units in \({S_{0}}\) are stable, yet allows moderate variation in wording and coverage. Empirically, higher thresholds over-filter useful semantics, whereas lower thresholds tend to admit noisy or unstable semantics.

If \(C \ge \tau \), we regard \({S_{0}}\) as stable and reliable, and it is directly used as the final semantic guidance. Otherwise, the semantic calibration unit is triggered to selectively aggregate the most consistent semantic units across \({S_{0}}\) and \(\{ S'\} \), yielding the calibrated final semantics \({S_{final}}\). This design enables \textit{Multisage} to explicitly quantify semantic reliability and suppress spurious artifacts introduced during augmentation, thereby providing the translation model with semantically grounded and cross-implementation–consistent guidance.

\textbf{Semantic Calibration Unit}. When the consistency score \(C\) is below the reliability threshold \(\tau \), the initial augmented semantics \({S_{0}}\) may contain unstable or noisy components. In this case, \textit{Multisage} activates the Semantic Calibration Unit, which aggregates only the semantic units that obtain majority support across the initial and mutant semantics, and reconstructs a calibrated final semantic set \({S_{final}}\).

We first construct the candidate semantic unit pool by collecting semantic units from both the initial semantics and the mutant semantics:

\begin{equation}
    \label{eq13}
U^{\text{all}}
=
U(S_{0})
\cup
\bigcup_{S' \in \{S'\}} U(S')
\end{equation}

For each unit \(u \in {U^{{\rm{all}}}}\), its support is computed as equation(\ref{eq13}). A semantic unit is considered reliable if it satisfies the majority-support condition \({\rm{supp}}(u) \ge \rho \).

We then define the consensus semantic unit set as:

\begin{equation}
    \label{eq14}
U^{\mathrm{cons}}
=
\left\{
u \;\middle|\;
u \in U^{\text{all}}
\ \text{and }\
\mathrm{supp}(u)\ge\rho
\right\}
\end{equation}

Finally, the calibrated semantics are obtained by reconstructing the structured semantic representation from the consensus unit set:

\begin{equation}
    \label{eq15}
S_{\text{final}}
=
\mathrm{Aggregate}\!\left(U^{\mathrm{cons}}\right)
\end{equation}

where \(\mathrm{Aggregate}( \cdot )\) denotes the reconstruction of a structured semantic representation (including summaries, API-level semantics, and test-case constraints) from the consensus unit set. In practice, the aggregation preserves the task structure of \({S_{0}}\) (e.g., summary, API-level semantics and test constraints), while replacing or supplementing unstable components using majority-supported units from \(U^{\mathrm{cons}}\). This allows \textit{Multisage} to deliver stable, cross-implementation–consistent, and noise-suppressed semantic guidance to the translation model, even when the initial semantics are unreliable.

The semantic consistency calibration module involves three threshold-based parameters: the unit-level similarity threshold $\delta$, the majority-support ratio $\rho$, and the global consistency threshold $\tau$. These parameters control how strictly \textit{Multisage} filters unstable or contradictory semantics. 

In this work, we set the default values to $\delta=0.7$, $\rho=0.5$, and $\tau=0.6$. These choices reflect intuitive design considerations: $\delta=0.7$ enforces a relatively high but not prohibitive similarity requirement at the semantic-unit level, $\rho=0.5$ requires at least majority agreement across variants, and $\tau=0.6$ ensures that global semantic instability triggers calibration only when sufficiently strong evidence accumulates. 

We stress that these values are not the result of fine-tuning for peak performance. As shown later in Section \ref{rq5}, \textit{Multisage} remains stable across wide parameter ranges, indicating that our framework is robust to reasonable variations in these thresholds.

\section{Experimental Setup}
\label{Experimental Setup}

\subsection{Datasets}
\textbf{Seed Code–Text Datasets}. We leverage two publicly available code–text datasets, XCodeEval \cite{xcodeeval} and XLCoST \cite{xlcost}, as the seed corpora for multi-semantic data construction. Both datasets provide executable source code paired with natural-language comments or summaries, enabling alignment between program logic and textual semantics.

XCodeEval is a large-scale multilingual benchmark derived from the Codeforces platform\footnote{https://codeforces.com/}, containing approximately 7,514 algorithmic problems with diverse executable contexts. XLCoST is collected from GeeksForGeeks\footnote{https://www.geeksforgeeks.org/} and provides paired code and natural-language descriptions across multiple programming languages such as C++ and Java.

Table \ref{tab:table1} summarizes the statistics of the extracted C++ samples. Since our study focuses on C++→Java translation, we extract the corresponding C++ programs and apply standard preprocessing, including comment removal and dataset deduplication.

\begin{table}[!t]
\centering
\renewcommand{\arraystretch}{1.5}
\caption{Statistics of the C++ samples in the seed code-text datasets.}
\label{tab:table1}
\begin{tabular}{C{3cm}C{3cm}}
\hline
\textbf{Source Dataset} & \textbf{\# C++ Samples} \\
\hline
XCodeEval & 95,425 \\
XLCoST    & 11,198 \\
\hline
\textbf{Total} & \textbf{106,623} \\
\hline
\end{tabular}
\end{table}

\textbf{Multi-Semantic Augmented Dataset}. Leveraging the proposed Multi-Semantic Data Construction pipeline, we systematically augment the original XCodeEval and XLCoST datasets. For each source instance, we generate 2–3 multi-semantic augmented samples, resulting in a large-scale dataset comprising 309,212 C++ samples and 311,890 Java samples.

\textbf{Test Dataset}. We evaluate our approach on the cross-lingual version of the HumanEval-X \cite{humaneval-x} benchmark suite, which consists of 164 distinct programming challenges. Each task is accompanied by a validated unit test harness and a reference implementation, enabling rigorous in terms of functional correctness and semantic consistency. Although HumanEval-X originates from code generation benchmarks, its cross-lingual setting provides functionally equivalent implementations and executable test suites, enabling reliable evaluation of semantic equivalence in code translation.

\subsection{Baselines}
\textbf{LLM baselines}. To investigate the performance boost provided by \textit{Multisage} across various LLM parameter scales, we selected models spanning the current mainstream parameter ranges. Specifically, we chose DeepSeek-V3 \cite{deepseekv3}, GPT-4 Turbo \cite{gpt4}, and Qwen-Plus \cite{qwen25} to represent HP-LLMs. Qwen2.5-Coder-Instruct-33B \cite{qwen25coder} and StarCoder2-15B \cite{starcoder2} were selected as MS-LLM baselines. Finally, StarCoder2-7B and Qwen2.5-Coder-7B serve as our baselines for $\ell$LMs. This selection enables a systematic evaluation of \textit{Multisage}'s generalization capability across different model scales.

\textbf{Specialized code translation baselines}. To compare \textit{Multisage} with established code translation techniques, we select four representative methods covering different translation paradigms:
(1) TransCoder \cite{transcoder}, which represents a Transformer-based neural machine translation approach;
(2) DOBF \cite{dobf}, a deobfuscation-based structured translation method designed to preserve program semantics and structure;
(3) TransCoder-ST \cite{transcoder-st}, which incorporates structured information such as AST to guide the translation process;
(4) TransCoder-IR \cite{transcoder-ir}, which employs an intermediate representation as a translation bridge to enhance semantic preservation; and (5) INTERTRANS \cite{intertrans}, which introduces intermediate languages as transitive bridges to facilitate cross-language translation by decomposing the translation process into multi-step transformations, and is implemented on top of the StarCoder2-15B model.

\textbf{Semantic Augmentation Capability Baselines}. 
To assess the effectiveness of the proposed Multi-Semantic Augmentation Module (M2) and examine the role of specialized augmentation models, we select state-of-the-art HP-LLMs as auxiliary baselines, including DeepSeek-V3, GPT-4 Turbo, and Qwen-Plus. We evaluate the ability of these general-purpose LLMs to directly generate semantic constraints, such as code summaries and unit tests, under zero-shot or few-shot prompting. Their generated constraints are then compared with those produced by the specialized M2 model in terms of accuracy, executability, and robustness. These baselines are introduced to support a dedicated evaluation of semantic augmentation quality in subsequent experiments.

\subsection{Evaluation Metrics}
\textbf{Code Translation Success Rate}. The Code Translation Success Rate measures the correctness of translated code in terms of executability and functional equivalence. We perform rigorous functional verification using the integrated unit test suite provided by the HumanEval-X benchmark. A translation is considered successful only if the generated target code passes all corresponding test cases, indicating that its functional behavior is consistent with the reference implementation.

Given that the primary objective of code translation is to ensure executability and functional correctness, we do not treat purely static similarity-based metrics as the main evaluation criterion. Prior studies have shown that relying solely on static similarity measures can be misleading, as models may achieve high scores while producing code that fails to compile or execute correctly. Accordingly, we adopt execution-based functional validation as the primary metric, and only incorporate structure- and data-flow–aware static metrics as complementary analysis tools.

\textbf{CodeBLEU}. CodeBLEU \cite{codebleu} is a structure-aware evaluation metric designed for code translation tasks and is used to complement execution-based validation by characterizing semantic consistency at the implementation level. Unlike static metrics that rely solely on n-gram matching, CodeBLEU extends BLEU with AST matching and data-flow analysis, enabling it to partially capture consistency in structural organization, variable dependencies, and data propagation patterns between the source and translated code. In particular, the data-flow component can be regarded as an approximation of semantic consistency at the implementation level. It should be emphasized that CodeBLEU does not guarantee functional equivalence, but instead provides complementary insights into the structural and semantic alignment of translation results beyond execution outcomes.

\subsection{Implementation Details}
\textit{Multisage} is primarily implemented in Python, utilizing the PyTorch deep learning framework and Hugging Face's Transformer library for model construction and management.

\textbf{Multi-Semantic Data Augmentation}. In the data construction phase of the Multi-Semantic Augmentation Module, we employ GPT-4 Turbo as a semantic generation agent to produce high-level semantic constraints, including unit tests and API information. This agent operates via carefully engineered prompts and an execution validation feedback loop to ensure the quality and diversity of the generated semantic constraints.

\textbf{Multi-Semantic Augmentation Model Training}. We fine-tuned the Multi-Semantic Augmentation Model based on the configurations derived from \cite{MFTCoder} to facilitate multi-task learning. During the training process, we employed the Adam optimizer, setting the per-GPU batch size to $2$, and aggregating the global batch size to $128$. For the learning rate schedule, we set the initial learning rate to $2 \times 10^{-4}$ and used a cosine annealing strategy with a warmup ratio of $3 \times 10^{-2}$, with the minimum learning rate set at $1 \times 10^{-5}$. For the fine-tuning strategy, we utilized the QLoRA INT4 quantization mode proposed in \cite{MFTCoder} and maintained a consistent fine-tuning parameter ratio of $2.52\%$ with the original work. Furthermore, to ensure optimal convergence and generalization performance for each model, we implemented an Early Stopping strategy to determine the model's termination point. 

All experiments were conducted on a multi-GPU server equipped with NVIDIA A6000 GPUs. During inference, \textit{Multisage} operates without additional fine-tuning and relies on prompt-based semantic augmentation, with an optional lightweight repair step applied to failed translations. All experiments were executed with fixed random seeds to ensure reproducibility. 

\subsection{Research Question}
To evaluate the effectiveness of the \textit{Multisage}, we formulate the following research questions (RQs):

\textbf{RQ1: How does the \textit{Multisage} improve the performance of LLM-based code translation and mitigate intervenable errors?} 

We evaluate two usage modes: (1) direct augmentation, where \textit{Multisage} is applied during initial translation, and (2) repair-based augmentation, where additional semantic guidance is injected only for initially failed translations to assess recovery capability. Performance is measured in terms of functional correctness, CodeBLEU, and error-type distribution shifts.

\textbf{RQ2: How does \textit{Multisage} perform against state-of-the-art code translation baselines?}

We compare \textit{Multisage} with specialized translation models and instruction-tuned (IT) LLMs under the same benchmark and evaluation metrics.

\textbf{RQ3: What is the quality of the semantic augmentation generated by \textit{Multisage}?}

We assess semantic quality from three perspectives. 

(1) \textit{Intrinsic validity}: We adopt a back-translation protocol in which generated semantics are used to reconstruct target-language implementations. The evaluation is reported using the BT-Pass@k metric, defined as the probability of recovering a functionally correct implementation.

(2) \textit{Comparative efficacy}: we compare \textit{Multisage}-generated semantics with reasoning-based baselines such as CoT. 

(3) \textit{Generative quality}: we directly contrast the semantic representations produced by \textit{Multisage} and HP-LLMs under identical targets.

\textbf{RQ4: How do individual components of \textit{Multisage} contribute to the overall translation performance?}

To address this question, we performed an ablation study to analyze and quantify the contribution of each component to the code translation task. By selectively removing key modules and examining the resulting performance changes, we isolate the distinct value that each component adds to the final generation quality.

\textbf{RQ5: How sensitive is \textit{Multisage} to the threshold parameters in the semantic consistency calibration module?}

We analyze the sensitivity of three parameters, namely the similarity threshold $\delta$, the majority-support ratio $\rho$, and the global consistency threshold $\tau$, by varying each parameter independently while keeping the others fixed at their default values, and observe the resulting performance changes.

\section{Results and Analysis}
\label{RQ}
\subsection{RQ1: How does the \textit{Multisage} improve the performance of LLM-based code translation and mitigate intervenable errors?}

\begin{figure}[!t]
\centering
\includegraphics[width=3.5in]{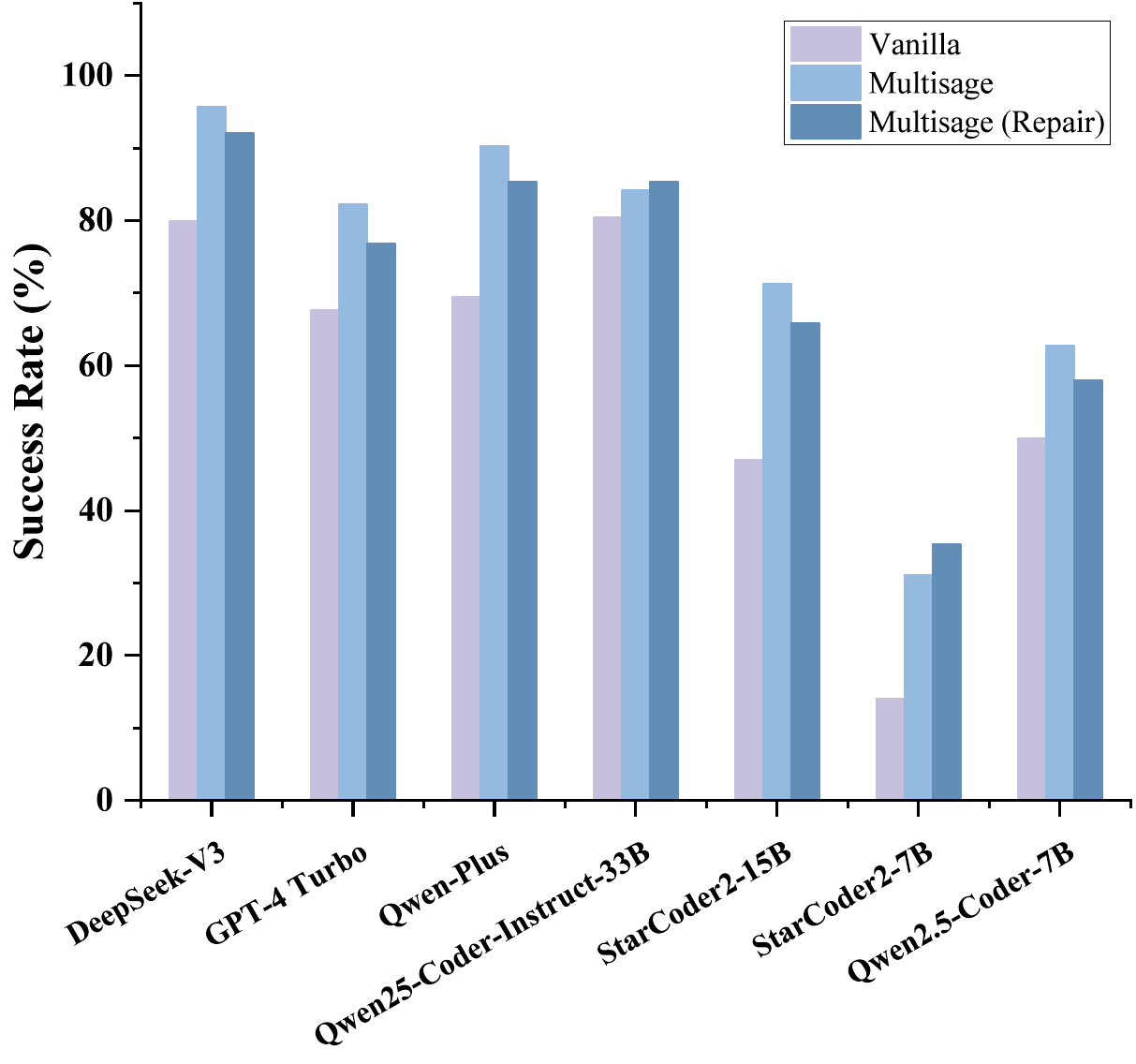}
\caption{Translation success rates under different Multisage settings.}
\label{fig_5_1}
\end{figure}

Fig. \ref{fig_5_1} and Table \ref{tab:rq1_1} report the translation performance under three settings: vanilla prompting, \textit{Multisage}, and \textit{Multisage} (Repair). Across all evaluated models, \textit{Multisage} consistently improves translation success rates over the vanilla setting. The gains are particularly pronounced for small and mid-scale models. For example, the success rate of StarCoder2-15B increases from 46.95\% to 71.34\%, while StarCoder2-7B improves from 14.02\% to 31.10\%.

In addition to higher execution success rates, \textit{Multisage} also improves CodeBLEU scores across all models, indicating better alignment in code structure and data-flow semantics. For instance, the CodeBLEU score of StarCoder2-15B increases from 30.86\% to 42.88\%. These results suggest that explicit semantic augmentation improves not only functional correctness but also the structural quality of generated code.

The \textit{Multisage} (Repair) setting provides an additional complementary mechanism. Instead of augmenting all translations, semantic guidance is selectively applied only to instances that fail during the initial generation. Although the overall success rate of \textit{Multisage} (Repair) does not always exceed direct \textit{Multisage}, it often achieves comparable or higher CodeBLEU scores. For weaker models, the repair setting can further improve the success rate beyond direct \textit{Multisage}. For example, StarCoder2-7B increases from 31.10\% under \textit{Multisage} to 35.37\% with \textit{Multisage} (Repair), indicating that targeted semantic guidance can recover a subset of previously failed translations.

To better understand how these improvements arise, we further analyze the distribution of translation errors before and after applying \textit{Multisage}. The detailed error statistics are reported in Appendix~\ref{appendixD}. Overall, errors related to dependency reasoning and data interpretation decrease substantially after semantic augmentation, particularly for small and mid-scale models. This observation suggests that many translation failures stem from missing or misinterpreted semantic constraints, which can be effectively mitigated through explicit semantic guidance. Meanwhile, model-specific errors remain relatively stable across settings, indicating that such failures are primarily determined by model capacity rather than the availability of semantic information.

\begin{table}[t]
\centering
\small
\renewcommand{\arraystretch}{1.18}
\setlength{\tabcolsep}{3pt}
\caption{Translation success rate and CodeBLEU under different settings.}
\label{tab:rq1_1}
\begin{adjustbox}{max width=\columnwidth}
\begin{tabular}{@{}c c c c@{}}
\toprule
\textbf{Model} & \textbf{Setting} & 
\makecell[c]{\textbf{Success}\\\textbf{Rate (\%)}} & 
\makecell[c]{\textbf{CodeBLEU}\\\textbf{(\%)}} \\
\midrule
\multirow{3}{*}{DeepSeek-V3}
& Vanilla            & 79.88 & 43.72 \\
& Multisage          & 95.73 & 52.35 \\
& Multisage (Repair) & 92.07 & 55.37 \\
\midrule
\multirow{3}{*}{GPT-4 Turbo}
& Vanilla            & 67.68 & 39.23 \\
& Multisage          & 82.32 & 47.42 \\
& Multisage (Repair) & 76.83 & 52.18 \\
\midrule
\multirow{3}{*}{Qwen-Plus}
& Vanilla            & 69.51 & 41.90 \\
& Multisage          & 90.24 & 48.19 \\
& Multisage (Repair) & 85.37 & 49.56 \\
\midrule
\multirow{3}{*}{\makecell[c]{Qwen2.5-Coder\\-Instruct-33B}}
& Vanilla            & 80.49 & 42.17 \\
& Multisage          & 84.15 & 53.60 \\
& Multisage (Repair) & 85.37 & 53.80 \\
\midrule
\multirow{3}{*}{StarCoder2-15B}
& Vanilla            & 46.95 & 30.86 \\
& Multisage          & 71.34 & 42.88 \\
& Multisage (Repair) & 65.85 & 43.25 \\
\midrule
\multirow{3}{*}{StarCoder2-7B}
& Vanilla            & 14.02 & 15.44 \\
& Multisage          & 31.10 & 30.24 \\
& Multisage (Repair) & 35.37 & 25.68 \\
\midrule
\multirow{3}{*}{\makecell[c]{Qwen2.5\\-Coder-7B}}
& Vanilla            & 50.00 & 28.39 \\
& Multisage          & 62.80 & 38.78 \\
& Multisage (Repair) & 57.93 & 38.43 \\
\bottomrule
\end{tabular}
\end{adjustbox}
\end{table}

\subsection{RQ2: How does \textit{Multisage} perform against state-of-the-art code translation baselines?}

\begin{table}[!t]
\centering
\renewcommand{\arraystretch}{1.4}
\caption{Performance comparison with state-of-the-art code translation baselines.}
\label{tab:rq2_1}
\begin{tabular}{l c c}
\hline
\textbf{Method (Scale)} & \textbf{Succ. (\%)} & \textbf{CodeBLEU (\%)} \\
\hline
TransCoder            & 21.06 & 61.24 \\
DOBF                  & 31.71 & 64.87 \\
TransCoder-ST         & 40.24 & 60.09 \\
TransCoder-IR         & 45.12 & 55.63 \\
INTERTRANS            & 67.34 & 69.47 \\
\hline
IT-LLM (7B Avg.)      & 38.11 & 28.79 \\
\textbf{Multisage (7B Avg.)} & \textbf{46.95} & \textbf{34.51} \\
\hline
IT-LLM (15B)          & 51.22 & 41.73 \\
\textbf{Multisage (15B)}     & \textbf{71.34} & \textbf{42.88} \\
\hline
\end{tabular}
\end{table}

Table \ref{tab:rq2_1} compares \textit{Multisage} with representative state-of-the-art code translation baselines under identical evaluation settings, including specialized neural translation models and instruction-tuned LLMs.

Compared with specialized translation models, \textit{Multisage} significantly improves execution success rates. In particular, \textit{Multisage} (15B) achieves 71.34\% success rate, surpassing the strongest baseline INTERTRANS (67.34\%) as well as earlier approaches such as TransCoder and DOBF by a large margin. These results indicate that semantic reliability modeling provides more effective guidance for preserving functional correctness than prior structural or intermediate-representation–based methods.

In terms of CodeBLEU, INTERTRANS achieves the highest score (69.47\%), suggesting that its multi-step translation process is effective in preserving surface-level structural similarity. In contrast, \textit{Multisage} achieves moderate but consistent improvements over instruction-tuned LLMs, increasing CodeBLEU from 28.79 to 34.51 at the 7B scale and from 41.73 to 42.88 at the 15B scale. This indicates that \textit{Multisage} improves structural alignment while primarily focusing on functional correctness.

When compared with instruction-tuned LLM baselines at the same model scale, \textit{Multisage} consistently improves both success rate and CodeBLEU. At the 7B scale, the success rate increases from 38.11\% to 46.95\%, while at the 15B scale it improves substantially from 51.22\% to 71.34\%. These results demonstrate that \textit{Multisage} enhances both functional correctness and semantic consistency beyond what can be achieved by instruction tuning alone.

Overall, the results highlight a clear distinction between structural similarity and functional correctness. While some methods (e.g., INTERTRANS) achieve higher CodeBLEU scores, \textit{Multisage} delivers superior execution performance, demonstrating the effectiveness of explicitly modeling semantic reliability in code translation.

\subsection{RQ3: What is the quality of the semantic augmentation generated by \textit{Multisage}?}

\textit{Intrinsic validity}. Table \ref{tab:rq3_1} reports the back-translation performance under different semantic input configurations. Results show that semantic representations with higher information density consistently lead to stronger intrinsic validity. For example, using code summaries alone substantially outperforms API-level comments, indicating that global semantic descriptions capture functional intent more effectively than localized interface documentation.

Adding executable constraints further improves semantic fidelity. When API-level test cases are combined with API comments, BT-Pass@1 increases from 62.94\% to 78.41\%, and BT-Pass@3 rises from 74.26\% to 88.69\%. An even larger improvement is observed when combining code summaries with function-level test cases, achieving the highest reconstruction success (93.05\% for BT-Pass@1 and 97.82\% for BT-Pass@3).

Moreover, the gap between BT-Pass@1 and BT-Pass@3 narrows as stronger semantic constraints are introduced, suggesting that richer semantic information reduces reliance on sampling diversity during reconstruction. These results confirm that \textit{Multisage} generates semantically faithful augmentations that provide effective functional constraints beyond simple executability.

\begin{table}[t]
\centering
\renewcommand{\arraystretch}{1.4}
\caption{Intrinsic validity evaluation using back-translation under different semantic configurations.}
\label{tab:rq3_1}
\begin{tabular}{C{5cm}C{3.5cm}C{3.5cm}}
\hline
\textbf{Semantic Type} & \textbf{BT-Pass@1 (\%)} & \textbf{BT-Pass@3 (\%)} \\
\hline
Code Summary & 82.53 & 89.17 \\
API Comments & 62.94 & 74.26 \\
API Comments + API Test Cases & 78.41 & 88.69 \\
Code Summary + Function Test Cases & 93.05 & 97.82 \\
\hline
\end{tabular}
\end{table}

\begin{figure}[!t]
\centering
\includegraphics[width=5in]{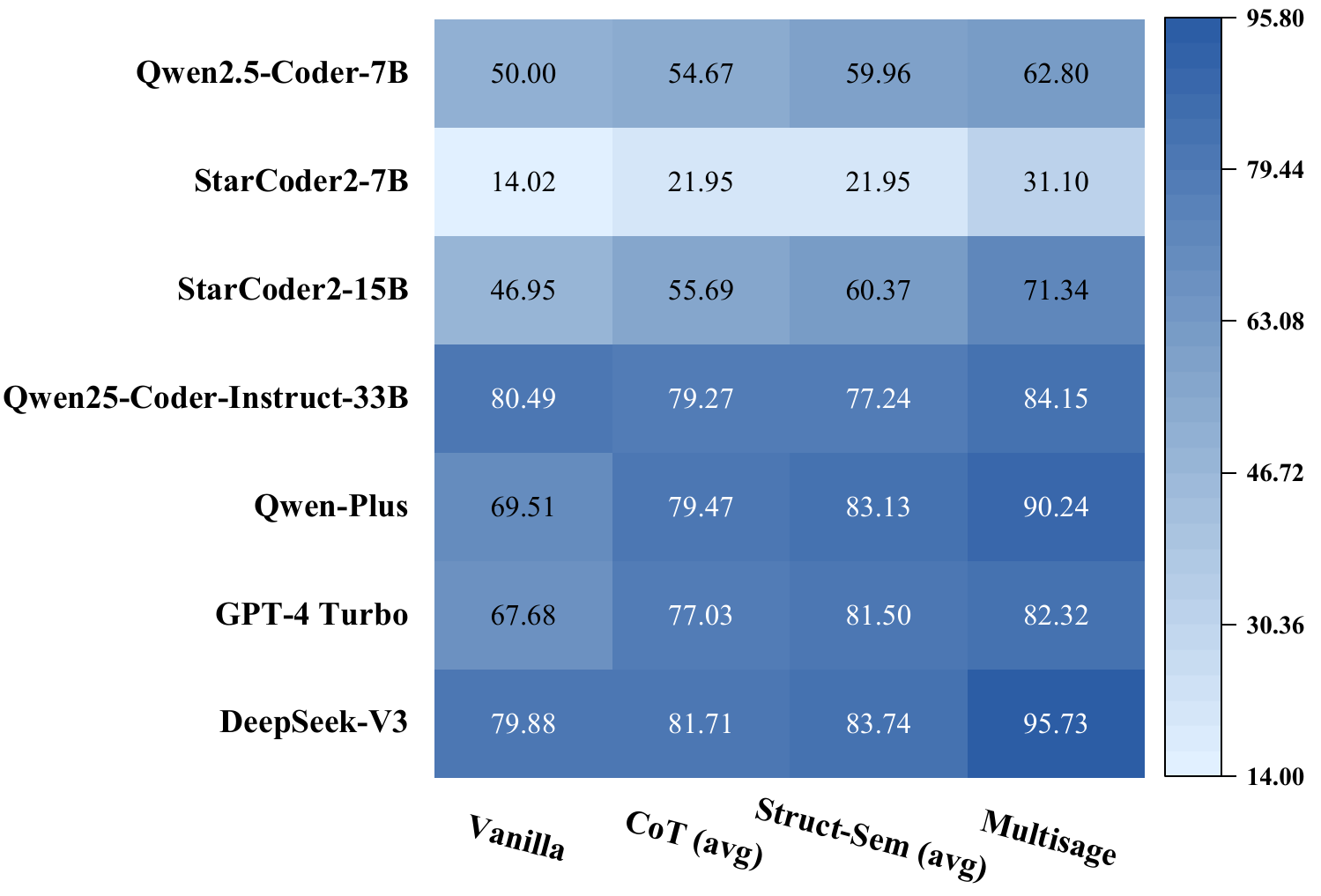}
\caption{Comparison of Translation Success Rates under Different Semantic Augmentation Strategies.}
\label{fig_5_3}
\end{figure}    

\begin{table*}[t]
\centering
\renewcommand{\arraystretch}{1.3}
\caption{CodeBLEU scores under different semantic augmentation strategies.}
\label{tab:rq3_2}
\begin{tabular}{C{4.5cm}C{2cm}C{2cm}C{3cm}C{2cm}}
\hline
\textbf{Model} & \textbf{Vanilla} & \textbf{CoT (avg)} & \textbf{Struct-Sem (avg)} & \textbf{Multisage} \\
\hline
DeepSeek-V3 & 43.72 & 47.33 & 50.68 & \textbf{52.35} \\
GPT-4 Turbo & 39.23 & 40.12 & \textbf{49.91} & \textbf{47.42} \\
Qwen-Plus & 41.90 & 43.57 & 47.84 & \textbf{48.19} \\
Qwen25-Coder-Instruct-33B & 42.17 & 49.26 & 52.09 & \textbf{53.60} \\
StarCoder2-15B & 30.86 & 28.45 & 40.77 & \textbf{42.88} \\
StarCoder2-7B & 15.44 & 12.15 & 28.97 & \textbf{30.24} \\
Qwen2.5-Coder-7B & 28.39 & 30.62 & 34.03 & \textbf{38.78} \\
\hline
\end{tabular}
\end{table*}

\textit{Comparative Efficacy}. Fig. \ref{fig_5_3} compares translation success rates under different semantic augmentation strategies. \textit{Multisage} consistently outperforms CoT prompting across all evaluated models. For example, DeepSeek-V3 improves from 81.71\% under CoT to 95.73\% with \textit{Multisage}, while StarCoder2-15B improves from 55.69\% to 71.34\%. Similar gains are observed for smaller models such as StarCoder2-7B (21.95\% vs. 31.10\%), indicating that explicit external semantics provide more effective guidance than internal reasoning traces, particularly when model capacity is limited.

This trend is also reflected in the CodeBLEU results reported in Table \ref{tab:rq3_2}. \textit{Multisage} achieves higher CodeBLEU scores than CoT across all evaluated models, indicating improved alignment in code structure and data-flow semantics. Together, these results demonstrate that \textit{Multisage} provides more effective semantic guidance than CoT, improving both functional correctness and implementation-level consistency.

\textit{Generative Quality}. Fig. \ref{fig_5_3} further compares \textit{Multisage} with a control setting in which HP-LLMs directly generate the same semantic descriptions used in \textit{Multisage} (denoted as Struct-Sem). \textit{Multisage} consistently achieves higher translation success rates across model scales. For example, DeepSeek-V3 improves from 83.74\% with Struct-Sem to 95.73\% with \textit{Multisage}, while StarCoder2-15B improves from 60.37\% to 71.34\%.

Table \ref{tab:rq3_2} provides complementary evidence on code quality. In most cases, \textit{Multisage} achieves higher CodeBLEU scores than Struct-Sem (e.g., DeepSeek-V3: 52.35 vs. 50.68; Qwen25-Coder-Instruct-33B: 53.60 vs. 52.09), indicating improved structural and data-flow alignment. An exception occurs for GPT-4 Turbo, where Struct-Sem obtains a slightly higher CodeBLEU score while \textit{Multisage} still achieves a higher execution success rate. This suggests that single-stage structural semantics may sometimes improve surface-level similarity, whereas \textit{Multisage} more reliably translates semantic guidance into execution-grounded correctness.

\subsection{RQ4: How do individual components of \textit{Multisage} contribute to the overall translation performance?}

\begin{table*}[t]
\centering
\renewcommand{\arraystretch}{1.4}
\caption{Overall ablation results of Multisage. Both translation success rate and CodeBLEU are reported in percentage. Values in parentheses indicate absolute percentage-point (pp) changes relative to the full Multisage setting.}
\label{tab:rq4_1}
\begin{adjustbox}{width=\textwidth}
\begin{tabular}{c|c|c|c c c c c c}
\hline
\toprule
\textbf{Model} & \textbf{Metric} & \textbf{Full} & \textbf{w/o M1} & \textbf{w/o Code Summary} & \textbf{w/o API Info.} & \textbf{w/o Test Cases} & \textbf{w/o Exec. Val.} & \textbf{w/o M3} \\
\midrule
\multirow{2}{*}{DeepSeek-V3}
 & Successful (\%) 
 & 95.73 
 & 89.02 (-6.71) 
 & 90.24 (-5.49) 
 & 79.88 (-15.85) 
 & 85.98 (-9.75) 
 & 84.15 (-11.58) 
 & 87.20 (-8.53) \\
 & CodeBLEU (\%) 
 & 52.35 
 & 45.92 (-6.43) 
 & 47.38 (-4.97) 
 & 48.96 (-3.39) 
 & 46.84 (-5.51) 
 & 51.21 (-1.14) 
 & 44.87 (-7.48) \\
\midrule

\multirow{2}{*}{GPT-4 Turbo}
 & Successful (\%) 
 & 82.32 
 & 76.83 (-5.49) 
 & 73.17 (-9.15) 
 & 60.37 (-21.95) 
 & 70.73 (-11.59) 
 & 70.12 (-12.20) 
 & 69.51 (-12.81) \\
 & CodeBLEU (\%) 
 & 47.42 
 & 41.08 (-6.34) 
 & 42.95 (-4.47) 
 & 44.61 (-2.81) 
 & 43.02 (-4.40) 
 & 46.37 (-1.05) 
 & 40.26 (-7.16) \\
\midrule
\multirow{2}{*}{Qwen-Plus}
 & Successful (\%) 
 & 90.24 
 & 80.49 (-9.75) 
 & 84.15 (-6.09) 
 & 70.73 (-19.51) 
 & 78.05 (-12.19) 
 & 81.10 (-9.14) 
 & 73.17 (-17.07) \\
 & CodeBLEU (\%) 
 & 48.19 
 & 42.11 (-6.08) 
 & 43.76 (-4.43) 
 & 45.28 (-2.91) 
 & 44.02 (-4.17) 
 & 47.02 (-1.17) 
 & 41.35 (-6.84) \\
\midrule
\multirow{2}{*}{Qwen25-Coder-Instruct-33B}
 & Successful (\%) 
 & 84.15 
 & 81.71 (-2.44) 
 & 76.22 (-7.93) 
 & 67.07 (-17.08) 
 & 71.34 (-12.81) 
 & 74.39 (-9.76) 
 & 68.90 (-15.25) \\
 & CodeBLEU (\%) 
 & 53.60 
 & 47.83 (-5.77) 
 & 49.62 (-3.98) 
 & 50.94 (-2.66) 
 & 49.18 (-4.42) 
 & 52.41 (-1.19) 
 & 46.92 (-6.68) \\
\midrule
\multirow{2}{*}{StarCoder2-15B}
 & Successful (\%) 
 & 71.34 
 & 69.51 (-1.83) 
 & 62.80 (-8.54) 
 & 48.78 (-22.56) 
 & 55.49 (-15.85) 
 & 62.20 (-9.14) 
 & 51.83 (-19.51) \\
 & CodeBLEU (\%) 
 & 42.88 
 & 36.94 (-5.94) 
 & 38.72 (-4.16) 
 & 40.05 (-2.83) 
 & 39.11 (-3.77) 
 & 41.83 (-1.05) 
 & 35.87 (-7.01) \\
\midrule
\multirow{2}{*}{StarCoder2-7B}
 & Successful (\%) 
 & 31.10 
 & 28.05 (-3.05) 
 & 21.34 (-9.76) 
 & 6.10 (-25.00) 
 & 14.63 (-16.47) 
 & 21.34 (-9.76) 
 & 8.54 (-22.56) \\
 & CodeBLEU (\%) 
 & 30.24 
 & 24.81 (-5.43) 
 & 26.03 (-4.21) 
 & 27.41 (-2.83) 
 & 25.96 (-4.28) 
 & 29.37 (-0.87) 
 & 23.95 (-6.29) \\
\midrule
\multirow{2}{*}{Qwen2.5-Coder-7B}
 & Successful (\%) 
 & 69.51 
 & 66.46 (-3.05) 
 & 60.37 (-9.14) 
 & 54.88 (-14.63) 
 & 56.71 (-12.80) 
 & 61.59 (-7.92) 
 & 54.88 (-14.63) \\
 & CodeBLEU (\%) 
 & 38.78 
 & 33.02 (-5.76) 
 & 34.69 (-4.09) 
 & 35.84 (-2.94) 
 & 34.21 (-4.57) 
 & 37.66 (-1.12) 
 & 32.41 (-6.37) \\
\midrule
\end{tabular}
\end{adjustbox}
\end{table*}

\begin{table*}[t]
\centering
\renewcommand{\arraystretch}{1.35}
\caption{Error-type distribution (\%) under different ablation settings. Error types A, B, and C correspond to dependency and logical errors, data parsing errors, and model-specific errors, respectively. Values in parentheses indicate absolute percentage-point (pp) changes relative to the full Multisage setting (positive means more errors).}
\label{tab:rq4_2}
\begin{adjustbox}{width=\textwidth}
\begin{tabular}{l|c|c|c c c c c c}
\toprule
\textbf{Model} & \textbf{Type} & \textbf{Full} & \textbf{w/o M1} & \textbf{w/o Code Summary} & \textbf{w/o API Info.} & \textbf{w/o Test Cases} & \textbf{w/o Exec. Val.} & \textbf{w/o M3} \\
\midrule
\multirow{3}{*}{DeepSeek-V3}
 & A (\%) & 3.05 & 7.93 (+4.88) & 5.49 (+2.44) & 15.24 (+12.19) & 11.59 (+8.54) & 12.20 (+9.15) & 9.76 (+6.71) \\
 & B (\%) & 1.22 & 2.44 (+1.22) & 3.66 (+2.44) & 4.88 (+3.66) & 2.44 (+1.22) & 3.66 (+2.44) & 2.44 (+1.22) \\
 & C (\%) & 0.00 & 0.61 (+0.61) & 0.61 (+0.61) & 0.00 (+0.00) & 0.00 (+0.00) & 0.00 (+0.00) & 0.61 (+0.61) \\
\midrule
\multirow{3}{*}{GPT-4 Turbo}
 & A (\%) & 12.80 & 15.85 (+3.05) & 17.68 (+4.88) & 26.83 (+14.03) & 20.73 (+7.93) & 20.12 (+7.32) & 20.12 (+7.32) \\
 & B (\%) & 4.88 & 6.10 (+1.22) & 8.54 (+3.66) & 10.98 (+6.10) & 7.93 (+3.05) & 9.76 (+4.88) & 9.15 (+4.27) \\
 & C (\%) & 0.00 & 1.22 (+1.22) & 0.61 (+0.61) & 1.83 (+1.83) & 0.61 (+0.61) & 0.00 (+0.00) & 1.22 (+1.22) \\
\midrule
\multirow{3}{*}{Qwen-Plus}
 & A (\%) & 7.32 & 15.85 (+8.53) & 10.37 (+3.05) & 20.73 (+13.41) & 12.80 (+5.48) & 11.59 (+4.27) & 17.68 (+10.36) \\
 & B (\%) & 2.44 & 3.05 (+0.61) & 5.49 (+3.05) & 7.32 (+4.88) & 8.54 (+6.10) & 6.71 (+4.27) & 7.93 (+5.49) \\
 & C (\%) & 0.00 & 0.61 (+0.61) & 0.00 (+0.00) & 1.22 (+1.22) & 0.61 (+0.61) & 0.61 (+0.61) & 1.22 (+1.22) \\
\midrule
\multirow{3}{*}{Qwen25-Coder-Instruct-33B}
 & A (\%) & 10.98 & 12.80 (+1.82) & 15.24 (+4.26) & 21.95 (+10.97) & 17.07 (+6.09) & 15.85 (+4.87) & 19.51 (+8.53) \\
 & B (\%) & 4.88 & 5.49 (+0.61) & 7.93 (+3.05) & 9.76 (+4.88) & 10.98 (+6.10) & 9.15 (+4.27) & 10.37 (+5.49) \\
 & C (\%) & 0.00 & 0.00 (+0.00) & 0.61 (+0.61) & 1.22 (+1.22) & 0.61 (+0.61) & 0.61 (+0.61) & 1.22 (+1.22) \\
\midrule
\multirow{3}{*}{StarCoder2-15B}
 & A (\%) & 21.34 & 21.95 (+0.61) & 25.61 (+4.27) & 36.59 (+15.25) & 28.66 (+7.32) & 24.39 (+3.05) & 33.54 (+12.20) \\
 & B (\%) & 3.66 & 4.88 (+1.22) & 7.32 (+3.66) & 8.54 (+4.88) & 10.98 (+7.32) & 9.15 (+5.49) & 9.15 (+5.49) \\
 & C (\%) & 3.66 & 3.66 (+0.00) & 4.27 (+0.61) & 6.10 (+2.44) & 4.88 (+1.22) & 4.27 (+0.61) & 5.49 (+1.83) \\
\midrule
\multirow{3}{*}{StarCoder2-7B}
 & A (\%) & 46.34 & 47.56 (+1.22) & 51.83 (+5.49) & 64.02 (+17.68) & 54.88 (+8.54) & 51.22 (+4.88) & 62.20 (+15.86) \\
 & B (\%) & 7.32 & 8.54 (+1.22) & 9.76 (+2.44) & 10.98 (+3.66) & 12.20 (+4.88) & 10.37 (+3.05) & 11.59 (+4.27) \\
 & C (\%) & 15.24 & 15.85 (+0.61) & 17.07 (+1.83) & 18.90 (+3.66) & 18.29 (+3.05) & 17.07 (+1.83) & 17.68 (+2.44) \\
\midrule
\multirow{3}{*}{Qwen2.5-Coder-7B}
 & A (\%) & 16.46 & 18.29 (+1.83) & 21.34 (+4.88) & 24.39 (+7.93) & 23.17 (+6.71) & 20.73 (+4.27) & 25.00 (+8.54) \\
 & B (\%) & 11.59 & 12.20 (+0.61) & 14.63 (+3.04) & 17.07 (+5.48) & 15.85 (+4.26) & 14.02 (+2.43) & 16.46 (+4.87) \\
 & C (\%) & 2.44 & 3.05 (+0.61) & 3.66 (+1.22) & 3.66 (+1.22) & 4.27 (+1.83) & 3.66 (+1.22) & 3.66 (+1.22) \\
\bottomrule
\end{tabular}
\end{adjustbox}
\end{table*}

To investigate the contribution of each component in \textit{Multisage}, we conduct an ablation study by removing individual modules while keeping all other components unchanged. Table~\ref{tab:rq4_1} reports the overall translation performance in terms of success rate and CodeBLEU, while Table~\ref{tab:rq4_2} summarizes the corresponding error-type distributions.

Across all evaluated models and settings, removing any component leads to a consistent degradation in performance. This observation indicates that \textit{Multisage} is not a loose combination of independent heuristics, but a tightly coupled framework in which different modules provide complementary capabilities.

\textit{code representation parsing module (M1)}.  
The code representation parsing module (M1) provides the structural foundation of \textit{Multisage} by reconstructing machine-interpretable semantic representations from source code. As shown in Table~\ref{tab:rq4_1}, removing M1 leads to a noticeable drop in both translation success rate and CodeBLEU. The error distribution in Table~\ref{tab:rq4_2} further shows an increase in dependency-related and logical errors. Without explicit structural abstractions, models are more likely to generate translations with broken semantic dependencies or inconsistent control logic. This effect is particularly pronounced for smaller and medium-sized models, highlighting the importance of structural guidance when model reasoning capacity is limited.

\textit{Semantic augmentation signals}.  
The semantic augmentation stage introduces complementary semantic views, including code summaries, API-level semantics, and test-case constraints. These signals jointly provide global and fine-grained semantic guidance during translation.

Removing code summaries results in a moderate but consistent decline in translation success rate, accompanied by more frequent global semantic mismatches such as incorrect algorithmic intent or misplaced control logic. This indicates that summaries serve as global semantic anchors that help maintain coherence throughout the translation.

API-level semantic information further contributes to preserving fine-grained dependencies. When API semantics are removed, Table~\ref{tab:rq4_1} shows substantial degradation in both success rate and CodeBLEU, while Table~\ref{tab:rq4_2} reveals a sharp increase in dependency-related errors. Without explicit API semantics, models struggle to infer external dependencies and usage contracts correctly.

Test-case-based semantic constraints provide direct executable supervision. Removing test cases leads to one of the largest drops in translation success rate, even when CodeBLEU decreases only moderately. The error analysis confirms that the absence of test cases significantly increases data parsing errors and logical inconsistencies, indicating that structural similarity alone cannot guarantee functional correctness.

\textit{Semantic consistency calibration module (M3)}.  
The execution validator and the semantic consistency calibration module jointly ensure the reliability of semantic guidance. The execution validator enforces executability by filtering invalid semantic samples, while M3 performs cross-view consistency calibration to suppress unreliable or contradictory semantics.

When these mechanisms are removed, Tables~\ref{tab:rq4_1} and~\ref{tab:rq4_2} show consistent performance degradation and a higher proportion of dependency-related and functional errors. This indicates that the main role of semantic calibration is not to increase the amount of semantic information, but to stabilize it by filtering noisy signals and reinforcing semantics that remain consistent across multiple views.

Overall, the ablation results demonstrate that \textit{Multisage} derives its effectiveness from the coordinated interaction between semantic construction and semantic calibration. M1 provides structural grounding, semantic augmentation modules supply complementary semantic constraints, and M3 ensures the reliability of these signals through cross-view consistency verification.

\subsection{RQ5: How sensitive is \textit{Multisage} to the threshold parameters in the semantic consistency calibration module?}
\label{rq5}
\begin{table}[t]
  \caption{Sensitivity of Multisage to the similarity threshold $\delta$.}
  \label{tab:rq5_1}
  \centering
  \begin{tabular}{ccc}
    \hline
    $\delta$ & Success Rate (\%) & CodeBLEU (\%) \\
    \hline
    0.5 & 83.24 & 45.83 \\
    0.6 & 86.57 & 46.92 \\
    \textbf{0.7 (default)} & \textbf{90.24} & \textbf{48.19} \\
    0.8 & 88.73 & 47.95 \\
    0.9 & 78.16 & 46.58 \\
    \hline
  \end{tabular}
\end{table}

\begin{table}[t]
  \caption{Sensitivity of Multisage to the majority-support ratio $\rho$.}
  \label{tab:rq5_2}
  \centering
  \begin{tabular}{ccc}
    \hline
    $\rho$ & Success Rate (\%) & CodeBLEU (\%) \\
    \hline
    0.3 & 81.42 & 46.38 \\
    0.4 & 84.65 & 47.53 \\
    \textbf{0.5 (default)} & \textbf{90.24} & \textbf{48.19} \\
    0.6 & 88.81 & 48.02 \\
    0.7 & 87.13 & 47.64 \\
    0.8 & 87.95 & 46.91 \\
    \hline
  \end{tabular}
\end{table}

\begin{table}[t]
  \caption{Sensitivity of Multisage to the global consistency threshold $\tau$.}
  \label{tab:rq5_3}
  \centering
  \begin{tabular}{ccc}
    \hline
    $\tau$ & Success Rate (\%) & CodeBLEU (\%) \\
    \hline
    0.4 & 88.71 & 46.82 \\
    0.5 & 89.63 & 47.65 \\
    \textbf{0.6 (default)} & \textbf{90.24} & \textbf{48.19} \\
    0.7 & 89.92 & 47.93 \\
    0.8 & 88.85 & 46.78 \\
    \hline
  \end{tabular}
\end{table}

Table~\ref{tab:rq5_1}, Table~\ref{tab:rq5_2}, and Table~\ref{tab:rq5_3} report the sensitivity of \textit{Multisage} to the three threshold parameters in the semantic consistency calibration module. We use Qwen-Plus as the representative backbone and adopt a one-factor-at-a-time design, sweeping each parameter while keeping the others fixed at their default values. Performance is evaluated using translation success rate and CodeBLEU on HumanEval-X.

\textit{Similarity threshold $\delta$}. As shown in Table~\ref{tab:rq5_1}, increasing $\delta$ from 0.5 to 0.7 improves translation performance, with the success rate rising from 83.24\% to 90.24\% and CodeBLEU increasing from 45.83\% to 48.19\%. However, when $\delta$ becomes overly strict ($\delta=0.9$), the success rate drops to 78.16\%, as many paraphrased but semantically correct units are discarded. Overall, \textit{Multisage} remains stable within $\delta\in[0.6,0.8]$, and the default value $\delta=0.7$ lies near the center of this stable region.

\textit{Majority-support ratio $\rho$}. Table~\ref{tab:rq5_2} shows a similar trend. Increasing $\rho$ from 0.3 to 0.5 raises the success rate from 81.42\% to 90.24\%, indicating that moderate cross-variant agreement effectively suppresses unreliable semantics. Further increasing $\rho$ leads to gradual performance degradation (87.95\% at $\rho=0.8$), as overly strict voting thresholds begin to remove useful semantic units. Nevertheless, performance remains relatively stable within $\rho\in[0.4,0.7]$.

\textit{Global consistency threshold $\tau$}. The sensitivity to $\tau$ is comparatively small. As reported in Table~\ref{tab:rq5_3}, the success rate varies only between 88.71\% and 90.24\% when $\tau$ ranges from 0.4 to 0.8, with CodeBLEU following a similar trend. The best performance is observed at $\tau=0.6$, while other values cause only minor changes.

Overall, \textit{Multisage} maintains stable translation performance across broad parameter intervals: $\delta\in[0.6,0.8]$, $\rho\in[0.4,0.6]$, and $\tau\in[0.5,0.7]$. The default configuration used in our main experiments lies near the center of these stable regions, indicating that the semantic consistency mechanism is robust to reasonable threshold variations and does not rely on delicate hyperparameter tuning.

\section{Discussion}
\begin{table}[t]
\centering
\caption{Performance comparison across model scales and specialized baselines.}
\label{tab:discussion}
\renewcommand{\arraystretch}{1.2}
\setlength{\tabcolsep}{4pt}
\begin{tabular}{lcccc}
\hline
\textbf{Method} & \textbf{Succ. (\%)} & \textbf{CodeBLEU (\%)} \\
\hline

\multicolumn{3}{l}{\textit{Specialized Translation Models}} \\
TransCoder & 28.05 & 61.24 \\
DOBF & 31.71 & 64.87 \\
TransCoder-ST & 40.24 & 60.09 \\
TransCoder-IR & 45.12 & 55.63 \\
INTERTRANS & 67.07 & 34.51 \\

\hline
\multicolumn{3}{l}{\textit{LLMs (Average by Scale)}} \\
HP-LLMs (Baseline) & 83.13 & 47.75 \\
HP-LLMs (+Multisage) & \textbf{93.29} & \textbf{54.99} \\

MS-LLMs (Baseline) & 72.57 & 42.41 \\
MS-LLMs (+Multisage) & \textbf{82.01} & \textbf{50.95} \\

$\ell$LMs (Baseline) & 39.64 & 28.17 \\
$\ell$LMs (+Multisage) & \textbf{58.23} & \textbf{37.20} \\

\hline
\end{tabular}
\end{table}

To further evaluate the generalization of \textit{Multisage} beyond the primary experimental setting, we conduct additional experiments on Java-to-Python translation. This setting introduces a different challenge profile, where explicit type constraints available in the source language are no longer strictly enforced in the target language, making it easier for semantic inconsistencies to emerge during generation.

The results demonstrate that \textit{Multisage} consistently improves both execution success rate and CodeBLEU across all model scales. Notably, the improvements are more pronounced for lightweight models, where the success rate increases substantially, suggesting that semantic augmentation effectively compensates for limited model capacity. At the same time, even high-performance LLMs benefit from \textit{Multisage}, indicating that its contribution is not redundant but complementary to existing model capabilities.

An interesting observation is that the improvements in CodeBLEU are generally smaller than those in execution success rate. This highlights a fundamental distinction between structural similarity and functional correctness. While some approaches may achieve high CodeBLEU scores by preserving surface-level structure, \textit{Multisage} focuses on improving semantic reliability, leading to more functionally correct programs even when structural similarity does not increase proportionally.

Furthermore, the consistent performance gains across diverse model families, including proprietary models such as GPT-4 Turbo and DeepSeek-V3 as well as open-source code models like StarCoder2-7B/15B and QwenCoder2.5-Coder-7B, suggest that \textit{Multisage} is largely model-agnostic. This indicates that its effectiveness primarily stems from the semantic augmentation and consistency calibration mechanisms rather than dependence on specific architectures or training paradigms.

Overall, these findings suggest that explicitly modeling semantic consistency across multiple views provides a robust and generalizable strategy for improving code translation, particularly in scenarios where model capacity alone is insufficient to guarantee functional correctness.

Detailed per-model results are provided in Appendix~\ref{appendix:full_results}, demonstrating that the observed improvements are consistent across all models rather than driven by a small subset of cases.

\section{Threats to Validity}
\subsection{Internal Validity}

The effectiveness of the multi-task fine-tuning module may be influenced by the choice of backbone model. To assess this threat, we instantiated the fine-tuning module with multiple alternative backbones while keeping all other components unchanged. Detailed results are provided in Appendix~\ref{appendixE}. Although absolute performance varies across configurations, the relative improvement trends remain consistent, suggesting that \textit{Multisage} does not depend on a specific fine-tuning backbone.

Another potential threat concerns the sensitivity of the semantic consistency calibration thresholds. If performance gains were observed only under narrowly tuned parameter values, conclusions could be confounded by hyperparameter optimization. We therefore conducted a parameter sensitivity analysis (Section~\ref{rq5}) by varying the similarity threshold $\delta$, the majority-support ratio $\rho$, and the global consistency threshold $\tau$ over broad ranges. \textit{Multisage} maintains stable performance within reasonable intervals, and the default configuration lies within these stable regions. This indicates that the observed improvements are not driven by brittle threshold tuning.

\subsection{External Validity}

Our experiments focus on function-level code translation, enabling controlled assessment of semantic equivalence and functional correctness. Although real-world software systems may involve project-level dependencies and broader execution contexts, the semantic artifacts employed in \textit{Multisage}, such as structured summaries, API-level semantics, and executable tests, are not inherently restricted to function-level granularity.

Evaluation is conducted on the C++→Java language pair, a widely studied and practically relevant cross-language translation setting. The consistent improvements observed under this representative scenario provide empirical evidence for the applicability of \textit{Multisage} in cross-language code translation tasks.

Finally, we evaluate \textit{Multisage} across models of different parameter scales and architectural families. The consistent trends observed across these heterogeneous backbones suggest that the framework operates independently of specific model configurations, although continued evaluation on newly emerging architectures would further broaden empirical coverage.

\section{Related Work}

This section reviews prior work on LLM-based code translation and semantic-enhanced code intelligence. Existing approaches can be broadly grouped based on whether and how additional semantic information is incorporated into the translation process.

\subsection{End-to-End LLM Code Translation}

Recent advances in code translation are largely driven by LLMs, which treat programs as structured sequences and learn cross-language mappings in an end-to-end manner. Early pretrained models such as CodeBERT \cite{codebert} demonstrate strong code understanding capabilities through large-scale pretraining objectives, including masked language modeling and replaced token detection. More recent code-oriented LLMs, such as QwenCoder \cite{qwen25coder}, StarCoder \cite{starcoder2}, and DeepSeek-Coder \cite{dscoder}, further improve code generation and translation capabilities through scaling and data expansion, providing stronger backbones for downstream code translation methods.

Building upon these increasingly powerful models, neural code translation methods have been developed to explicitly model cross-language mapping. A key milestone in neural code translation is TransCoder \cite{transcoder}, which enables unsupervised cross-language transfer through denoising autoencoding and back-translation. Subsequent work further improves this paradigm by enhancing data quality and strengthening functional understanding. For example, deobfuscation-based objectives \cite{dobf} encourage models to recover semantic structure, while approaches such as Function-to-Style \cite{Function-to-Style} and related data augmentation methods improve translation performance through semantically enriched supervision.

Beyond improvements in model capacity and training objectives, more recent studies focus on improving the translation process itself. Reasoning-oriented approaches, such as EffiReasonTrans \cite{EffiReasonTrans}, introduce multi-step reasoning processes to enhance logical consistency. Other methods formulate translation as an iterative refinement process, including dialogue-based or multi-agent generation frameworks \cite{beyound_code_pairs}, which progressively improve intermediate outputs. 

Despite these advances, end-to-end neural translation models largely depend on implicit statistical correlations and lack mechanisms to explicitly represent, verify, and enforce program semantics during generation. As a result, they often produce outputs that are syntactically plausible yet semantically inconsistent, particularly in scenarios involving complex control flow, data dependencies, or API interactions.

\subsection{External-Information–Augmented Code Translation}

Another line of work attempts to improve code translation by incorporating external semantic or structural information. The key intuition is that exposing latent program semantics through program analysis artifacts or auxiliary supervision can help models better preserve functional intent.

Early approaches primarily rely on structured intermediate representations to encode program structure. For example, TreeBERT \cite{tree-tree} leverages ASTs to capture syntactic dependencies, while INTERTRANS \cite{intertrans} utilizes graph-based representations to model structural and semantic relationships across programs. Subsequent work further extends this line by exploring richer representation forms. SynCoBERT \cite{SynCoBERT} combines ASTs with code and comments as multimodal inputs, while other methods \cite{Syntax_and_Domain}, \cite{code_distillation} incorporate graph-based or representation learning techniques to better model program semantics.

Beyond structural representations, recent research introduces more diverse forms of semantic grounding. Some approaches adopt natural language as an intermediate representation. For instance, BabelCoder \cite{BabelCoder} generates natural language specifications to describe program behavior and guide translation and refinement. Other methods focus on structured decomposition and modular translation. Wang et al. \cite{program_skeletons} represent programs as high-level templates with placeholders linked to local semantic constraints, while AlphaTrans \cite{alphatrans} constructs target-language skeletons and performs compositional translation based on program decomposition and dependency analysis.

More recent work incorporates execution-level and system-level signals to further enhance semantic correctness. ExeCoder \cite{exectrans} introduces executability-oriented representations and leverages instruction tuning to improve runtime correctness. EvoC2Rust \cite{evoctrans} integrates static analysis, compiler feedback, and staged refinement in a system-level pipeline. Execution-guided methods, such as UniTrans \cite{unitrans} and related approaches, utilize automatically generated test cases or runtime feedback to iteratively repair translation errors. In addition, system-level frameworks incorporating API grounding and external constraints further improve translation reliability in complex real-world scenarios.

Despite these advances, existing approaches exhibit two fundamental limitations. First, many methods rely on external resources such as analyzers, documentation, test suites, or compiler feedback, which may be unavailable or costly in practical settings. Second, and more importantly, these approaches typically assume that the introduced semantic signals are reliable, without explicitly modeling their consistency or robustness across different representations. As a result, noisy, partial, or conflicting semantic information may still mislead the translation process.

In contrast, \textit{Multisage} constructs multiple complementary semantic views directly from source code and introduces a cross-view consistency mechanism to explicitly assess and calibrate their reliability. By identifying stable semantic signals across semantically equivalent variants, Multisage enables the model to leverage not only richer semantic information but also more trustworthy and robust guidance for code translation.

\section{Conclusion}

This paper proposes \textit{Multisage}, a multi-semantic augmentation and self-calibration framework for LLM-based code translation. The framework integrates three cooperative components: semantic representation parsing, multi-semantic augmentation, and semantic consistency calibration, to systematically construct diverse and complementary semantic information directly from source code without relying on external resources. By providing explicit and reliable semantic guidance, \textit{Multisage} alleviates the limitations of purely probabilistic translation pipelines and improves the functional correctness and reliability of cross-language code translation.

Experimental results demonstrate that \textit{Multisage} consistently improves translation performance across LLMs with different architectures and parameter scales. On the HumanEval-X benchmark, it achieves up to a 2.22× improvement in translation success rate compared with vanilla models. The framework also exhibits strong cross-model generalization, benefiting both general-purpose and code-specialized LLMs. These improvements arise from the richer and more reliable semantic constraints introduced by multi-semantic augmentation and consistency calibration. Furthermore, back-translation–based evaluations confirm that the semantics generated by \textit{Multisage} possess high intrinsic validity and can effectively support the reconstruction of functionally equivalent implementations.

In future work, we plan to further evaluate the robustness of \textit{Multisage} in more complex cross-language migration scenarios, such as library substitution, framework migration, and heterogeneous runtime environments. We also aim to more tightly integrate semantic augmentation with the translation process itself to enhance the practical reliability of LLM-based code translation in real-world software engineering environments.

\bibliographystyle{unsrt}  
\bibliography{references}

\appendix

\section{Prompt Templates for Multi-Semantic Data Construction}
\label{appendixA}
\subsection{Semantic Generation Prompt}
\begin{tcolorbox}[
    left=2pt,
    colback=white,                 
    colframe=black!70,             
    title=Semantic Generation Prompt,
    coltitle=white,                
    fonttitle=\bfseries,
    colbacktitle=black!70,         
    sharp corners,                 
    boxrule=0.8pt                  
]
\small
\begin{verbatim}
You are an expert [Lang] developer and 
software engineer. 
I will provide you with a [Lang] function and 
any external APIs it uses. 
Your task is to generate the following 
outputs:
1. Code Summary
2. Function Test Cases
3. External API Comments and Test Cases
[Code]
[API List]
\end{verbatim}
\end{tcolorbox}

\subsection{Semantic Refinement Prompt}

\begin{tcolorbox}[
    left=2pt,
    colback=white,                 
    colframe=black!70,             
    title=Semantic Refinement Prompt,
    coltitle=white,                
    fonttitle=\bfseries,
    colbacktitle=black!70,         
    sharp corners,                 
    boxrule=0.8pt                  
]
\small
\begin{verbatim}
You are an expert [Lang] developer and 
software engineer.
Based on the validation feedback, adjust the 
test cases for the function or specified 
API to resolve the reported issues.
[Target Function/API]
[Corrective Signal]
\end{verbatim}
\end{tcolorbox}

\section{Multi-Task Optimization Details of the Multi-Semantic Augmentation Model}
\label{appendixB}
\subsection{Token-Level Normalization Across Semantic Tasks}

Due to the heterogeneity of semantic generation tasks, different tasks exhibit substantial variation in dataset size and output sequence length. 
Directly aggregating task-level losses without normalization may introduce bias toward tasks with longer sequences or larger datasets. 

To mitigate this issue, we normalize the loss of each task by the number of valid target tokens. 
For task $T_i$ with dataset $D_i = \{(x_i^{(j)}, y_i^{(j)})\}_{j=1}^{M_i}$, the normalized loss is defined as:

\begin{equation}
\label{eqA1}
\mathcal{L}_i(\theta)
=
\frac{
\sum_{j=1}^{M_i}
\sum_{k=1}^{T_{ij}}
-\log \big(p_\theta(t_{ijk})\big)
}{
\sum_{j=1}^{M_i} T_{ij}
}
\end{equation}

where $T_{ij}$ denotes the number of valid tokens in the $j$-th sample, 
$t_{ijk}$ represents the $k$-th target token, and 
$p_\theta(t_{ijk})$ is the predicted probability under parameters $\theta$.

This normalization ensures that the contribution of each task reflects its intrinsic semantic complexity rather than its sequence length.

\subsection{Focal Adjustment for Task Difficulty}

Different semantic tasks vary in abstraction level and prediction difficulty. 
To prevent easy samples from dominating optimization, we introduce a focal-style adjustment mechanism that emphasizes relatively hard instances.

For task $T_i$, the focal-adjusted loss is defined as:

\begin{equation}
\label{eqA2}
\begin{aligned}
\mathcal{L}_i^{\text{focal}}(\theta)
&=
\frac{1}{M_i}
\sum_{j=1}^{M_i}
\alpha_i \cdot (1 - C_{ij})^\gamma \cdot Q_{ij}, \\
Q_{ij}
&=
- \frac{1}{T_{ij}}
\sum_k \log\big(p_\theta(t_{ijk})\big)
\end{aligned}
\end{equation}

where $C_{ij}$ denotes the mean token-level confidence of sample $j$, 
$\alpha_i$ is a task-level coefficient, and 
$\gamma$ controls the focusing strength.

This mechanism reduces the gradient contribution of high-confidence (easy) samples and increases emphasis on more challenging semantic predictions.

\subsection{Adaptive Task Weighting}

To coordinate convergence across heterogeneous semantic objectives, we adopt an adaptive weighting strategy inspired by FAMO.

At iteration $t$, the aggregated gradient is computed as:

\begin{equation}
\label{eqA3}
g_t
=
\sum_{i=1}^{N}
w_i^{(t)}
\nabla_{\theta_t} \mathcal{L}_i(\theta_t)
\end{equation}

with parameter update:

\begin{equation}
\label{eqA4}
\theta_{t+1}
=
\theta_t
-
\eta g_t
\end{equation}

Task weights are updated according to the relative improvement:

\begin{equation}
\label{eqA5}
\begin{aligned}
w_i^{(t+1)}
&=
\frac{c_i(\eta, g_t)}
{\sum_{j=1}^{N} c_j(\eta, g_t) + \varepsilon}, \\
c_i(\eta, g_t)
&=
\frac{
\mathcal{L}_i(\theta_t)
-
\mathcal{L}_i(\theta_t - \eta g_t)
}{
\mathcal{L}_i(\theta_t) + \varepsilon
}
\end{aligned}
\end{equation}

where $\varepsilon$ ensures numerical stability.
This update reflects the relative convergence speed of each task under the current optimization direction.

\subsection{Final Multi-Task Objective}

The final optimization objective integrates the normalized multi-task loss and the focal-adjusted component:

\begin{equation}
\label{eqA6}
\mathcal{L}_{\text{Multisage}}(\theta)
=
\lambda_1 \mathcal{L}(\theta)
+
\lambda_2
\sum_{i=1}^{N}
\mathcal{L}_i^{\text{focal}}(\theta)
\end{equation}

where $\lambda_1$ and $\lambda_2$ control the relative contribution of the base and focal terms.

To improve parameter efficiency, we employ PEFT~\cite{peft} to fine-tune a small subset of parameters while keeping the backbone largely frozen.

\subsection{Hyperparameter Settings}

Following~\cite{MFTCoder}, we set the focusing parameter to $\gamma = 2$ and the stability constant to $\varepsilon = 1\times10^{-6}$.
Unless otherwise specified, $\lambda_1 = 1$ and $\lambda_2 = 1$.

\subsection{Equivalence Mutator: Transformation Rules}

This section details the static semantics-preserving transformations used by the equivalence mutator.

\section{Equivalence Mutator: Transformation Rules}
\label{appendixC}
\subsubsection{Expression-Level Transformations}

Expression-level rewrites are applied to side-effect-free expressions. 
Typical transformation patterns include:

\begin{itemize}
    \item Commutative reordering: $a + b \leftrightarrow b + a$
    \item Neutral updates: $x = x + 0$, $x = x \times 1$
    \item Boolean equivalence rewrites (e.g., De Morgan’s laws)
\end{itemize}

These transformations are applied only when type consistency and absence of side effects are verified via static analysis.

\subsubsection{Control-Flow–Preserving Transformations}

Control-flow transformations modify the syntactic structure of branching and looping constructs while preserving execution semantics. 
Representative patterns include:

\begin{itemize}
    \item Negating branch predicates and swapping branch bodies
    \item Refactoring nested conditionals into equivalent cascades
    \item Rewriting conditional expressions into explicit if-else statements
\end{itemize}

All control-flow rewrites are performed under CFG constraints to ensure that reachable execution paths and guard conditions remain semantically equivalent.

\subsubsection{API-Level Equivalent Substitutions}

When multiple standard library functions provide equivalent functionality under identical preconditions, we substitute calls with verified alternatives. 
These substitutions are restricted to a curated set of API pairs that have been manually validated to preserve parameter and return-type contracts.

Mutants that fail compilation or execution checks are discarded during validation.

\section{Error Distribution Analysis}
\label{appendixD}

\begin{figure*}[!t]
\centering

\begin{minipage}{\textwidth}
\centering
\subfloat[]{\includegraphics[width=0.32\textwidth]{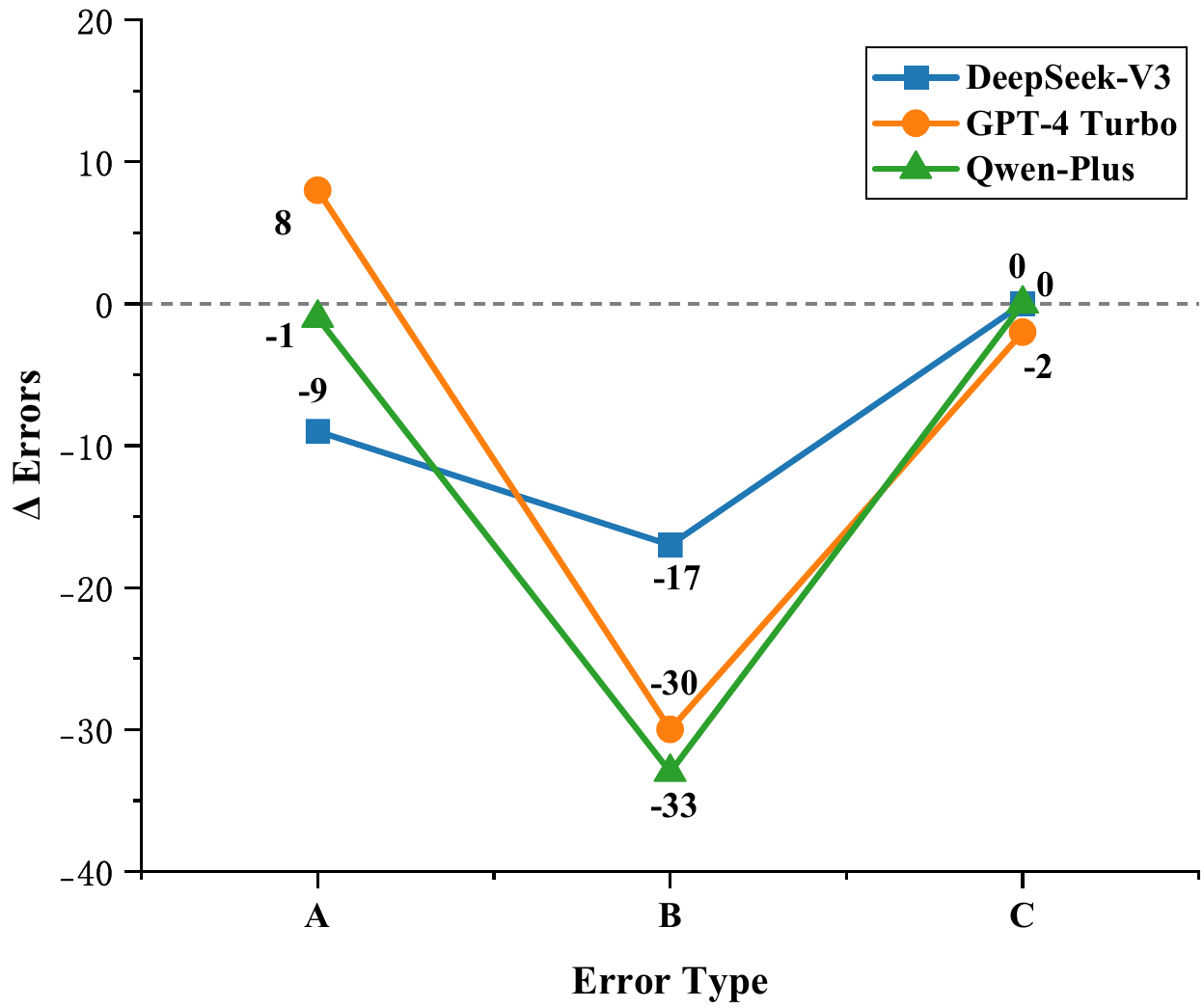}\label{fig_5_2_1}}
\hspace{2pt}
\subfloat[]{\includegraphics[width=0.32\textwidth]{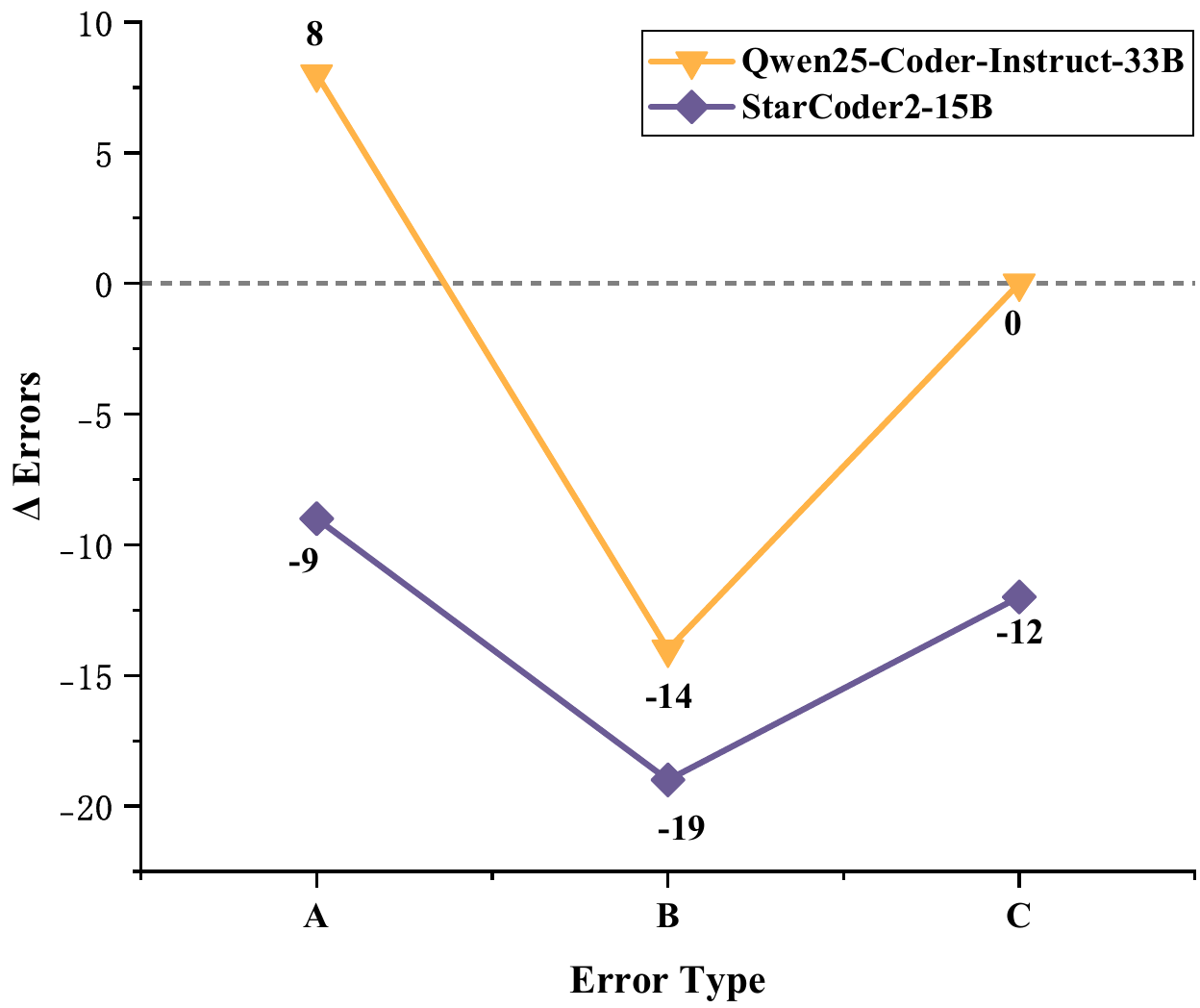}\label{fig_5_2_2}}
\hspace{2pt}
\subfloat[]{\includegraphics[width=0.32\textwidth]{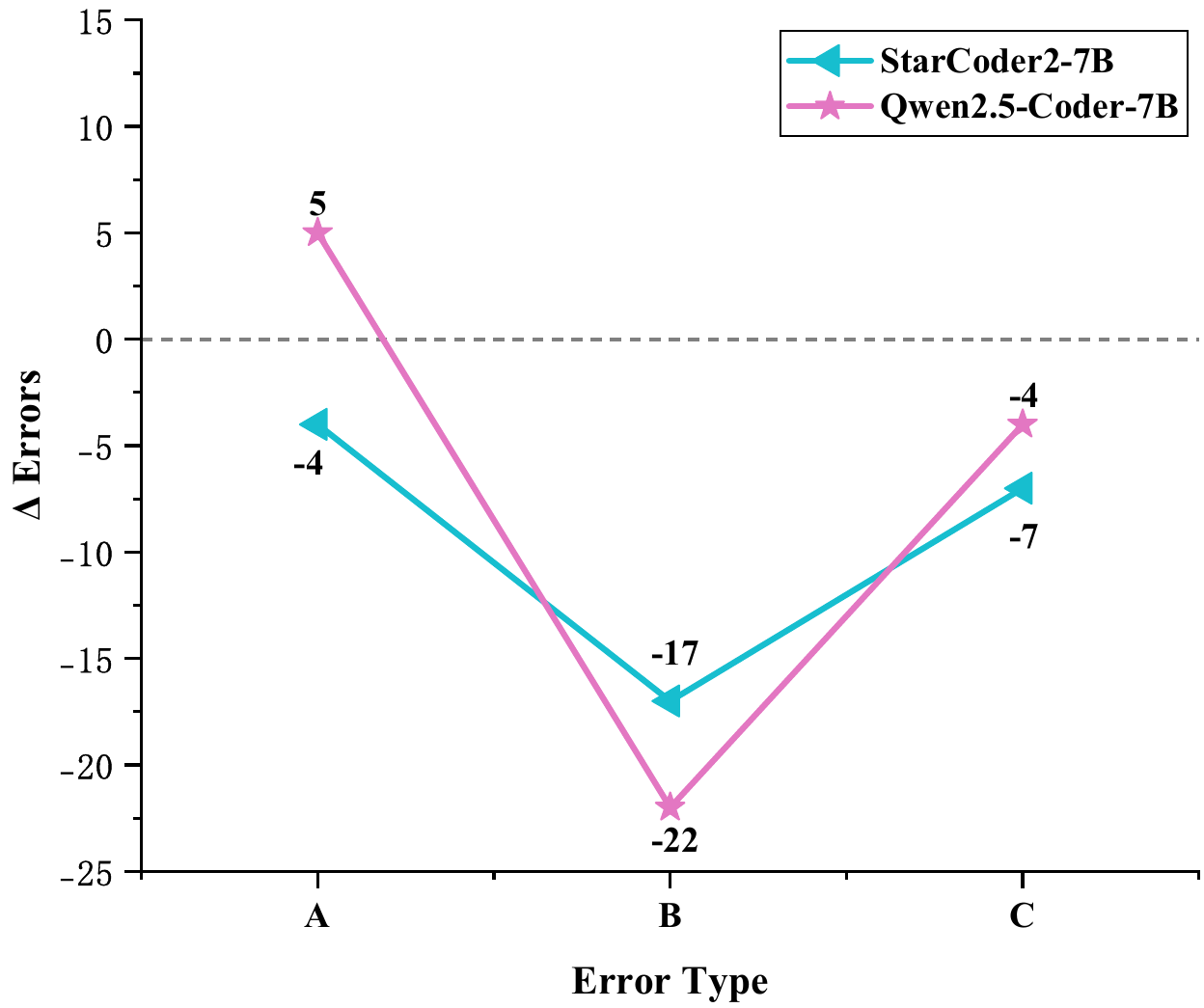}\label{fig_5_2_3}}
\end{minipage}

\vspace{-4pt} 

\begin{minipage}{\textwidth}
\centering
\subfloat[]{\includegraphics[width=0.32\textwidth]{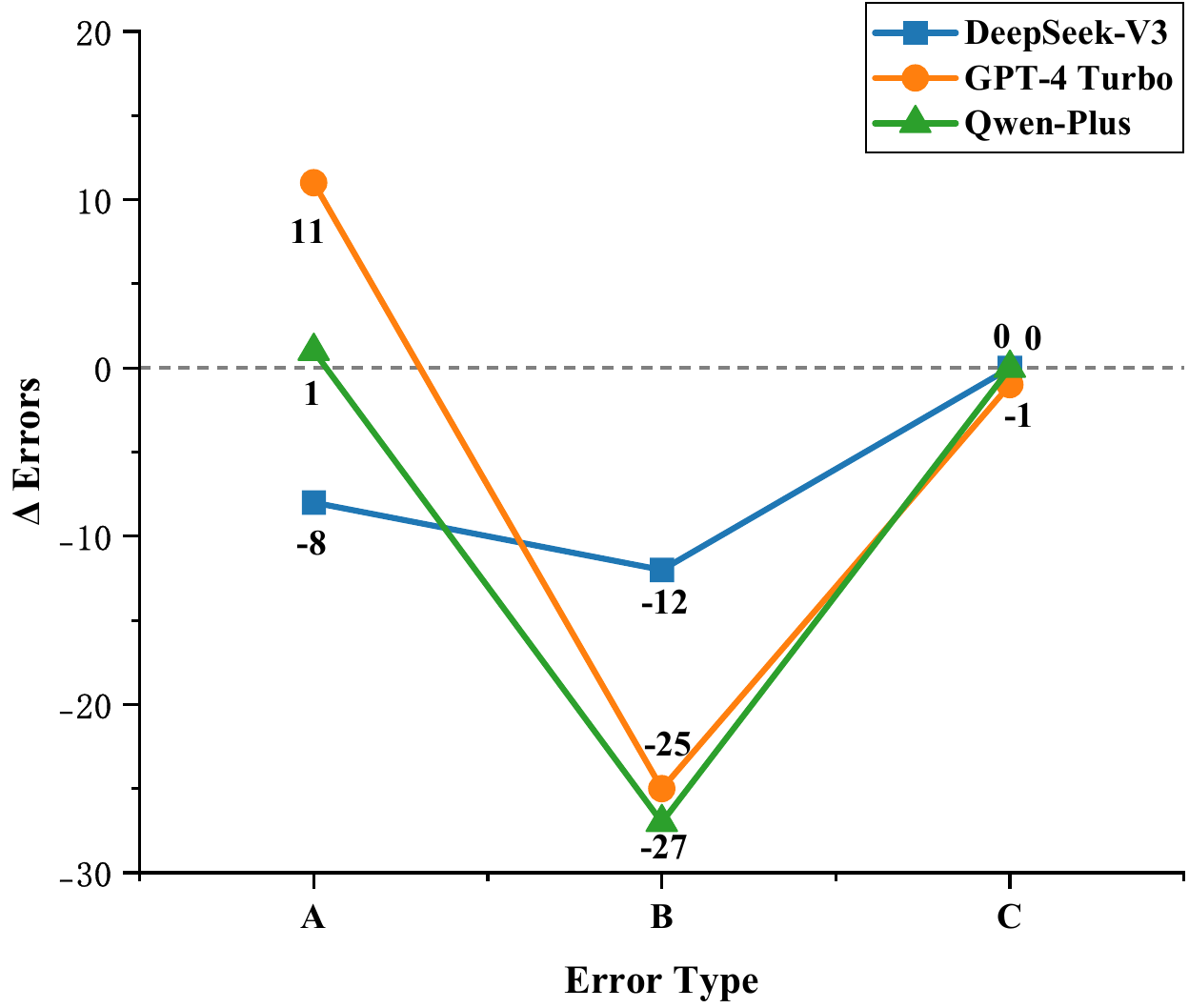}\label{fig_5_2_4}}
\hspace{2pt}
\subfloat[]{\includegraphics[width=0.32\textwidth]{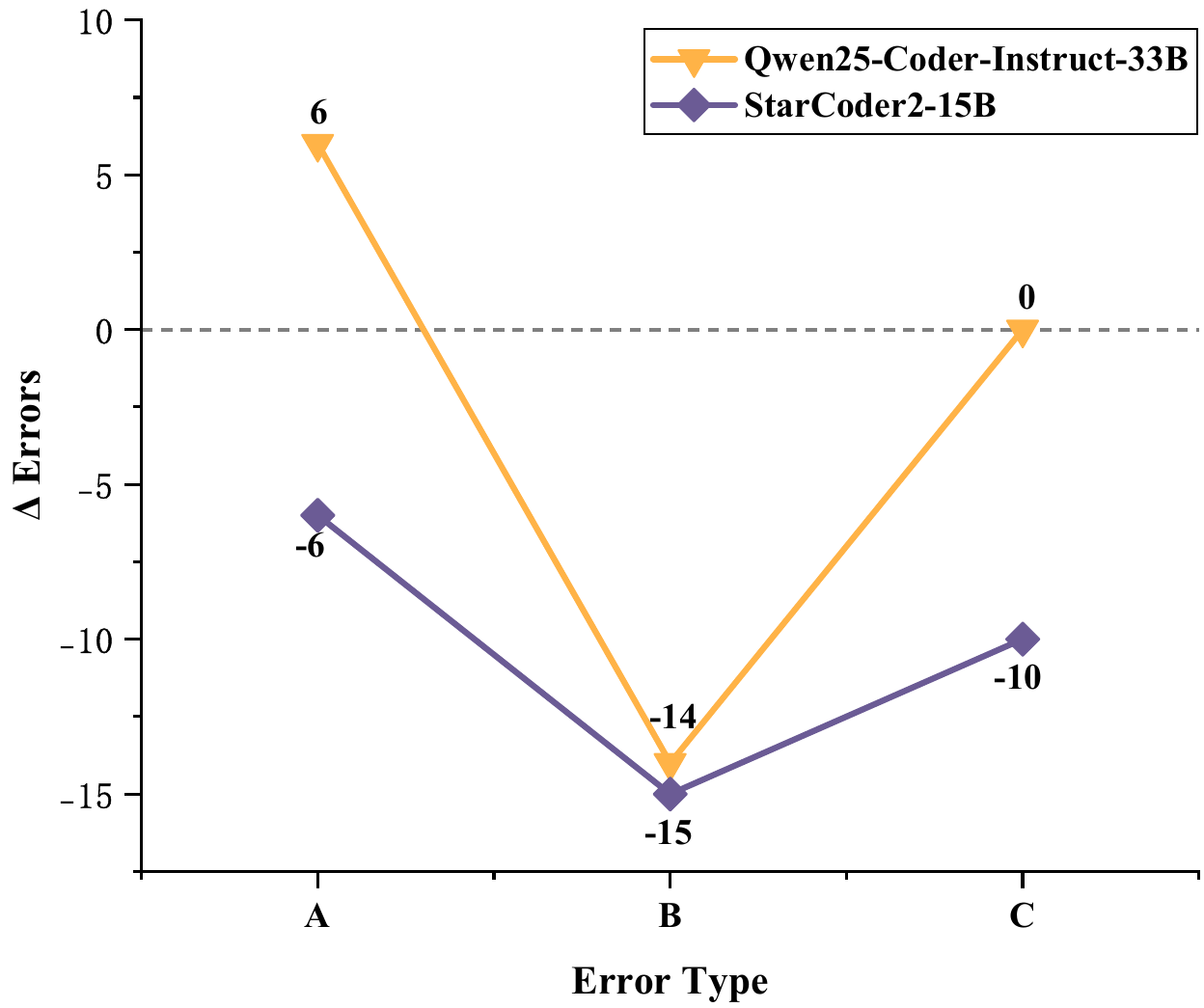}\label{fig_5_2_5}}
\hspace{2pt}
\subfloat[]{\includegraphics[width=0.32\textwidth]{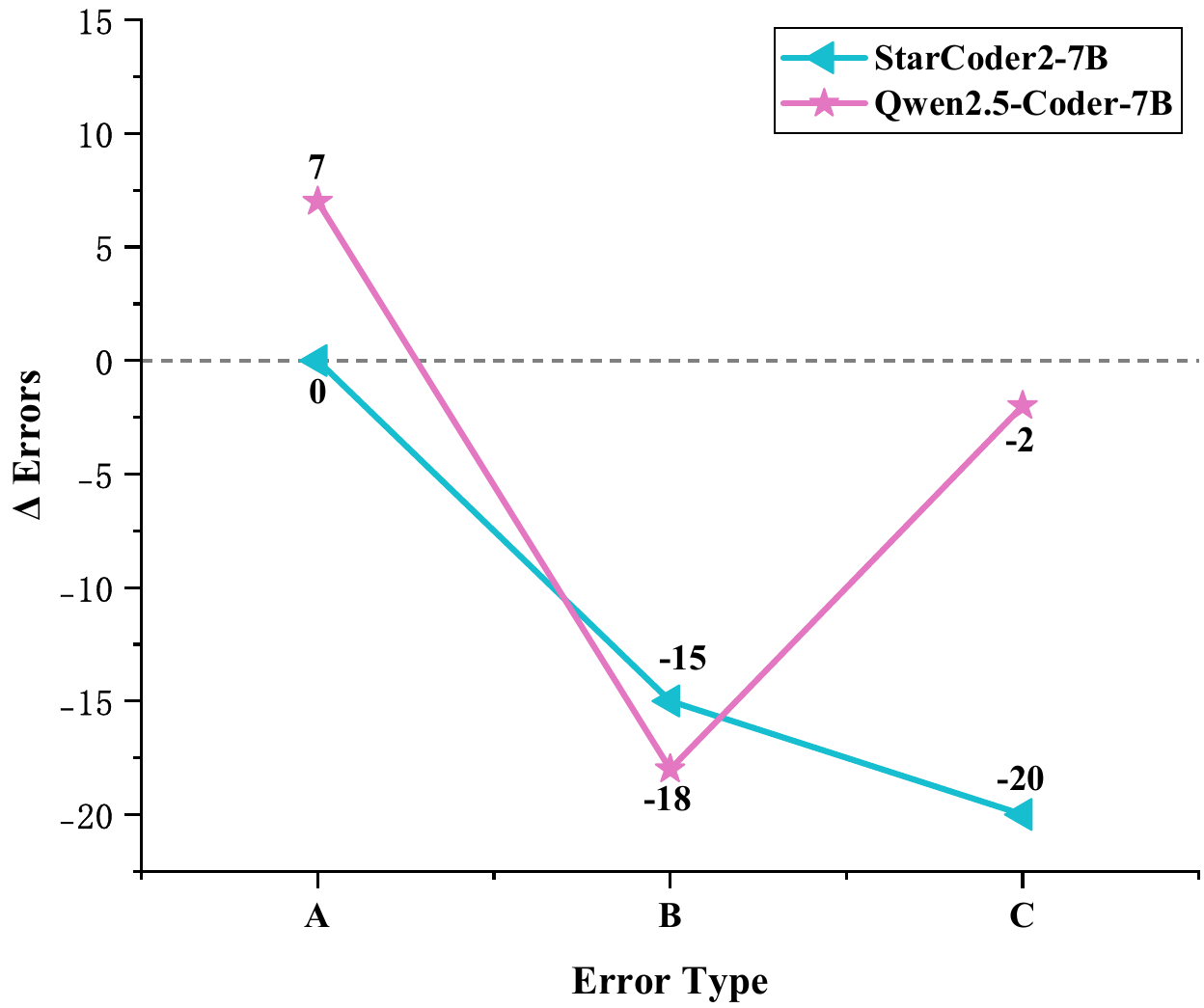}\label{fig_5_2_6}}
\end{minipage}

\caption{Error distribution shifts under Multisage and Multisage (Repair). Each error type corresponds to a representative category defined in the Motivation section.
(a) HP-LLMs with Multisage.
(b) MS-LLMs with Multisage.
(c) $\ell$LLMs with Multisage.
(d) HP-LLMs with Multisage (Repair).
(e) MS-LLMs with Multisage (Repair).
(f) $\ell$LLMs with Multisage (Repair).}
\label{fig_5_2}
\end{figure*}

To better understand the impact of semantic augmentation on translation failures, we analyze the distribution of error types before and after applying \textit{Multisage}. The results are shown in Fig. \ref{fig_5_2}. 

Overall, errors related to dependency reasoning and data interpretation decrease substantially after introducing semantic augmentation, particularly for small and mid-scale models. In contrast, model-specific errors remain relatively stable, suggesting that these failures are primarily determined by model capacity rather than missing semantic information. These observations further support the conclusion that a large portion of translation failures arises from missing or misinterpreted semantic constraints.

\section{Backbone Sensitivity Analysis}
\label{appendixE}

\begin{table*}[!t]
\centering
\renewcommand{\arraystretch}{1.25}
\caption{Translation success rates (\%) of Multisage under different fine-tuning backbones.
Rows denote the evaluated translation models, while columns correspond to the backbone models used to instantiate the multi-task augmentation module.}
\label{tab:internal_validity_backbone}
\begin{tabular}{c|cccc}
\hline
\multicolumn{1}{c|}{\multirow{2}{*}{\textbf{Translation Model}}} &
\multicolumn{4}{c}{\textbf{Fine-Tuning Backbone for the Multi-Task Augmentation Module}} \\
\cline{2-5}
\multicolumn{1}{c|}{} &
\textbf{CodeLlama-13B} & \textbf{StarCoder2-15B} & \textbf{CodeLlama-7B} & \textbf{Qwen2.5-Coder-7B} \\
\hline 
DeepSeek-V3 & 95.73 & 90.24 & 87.80 & 89.02 \\
GPT-4 Turbo & 82.32 & 75.61 & 73.17 & 74.39 \\
Qwen-Plus & 90.24 & 83.54 & 81.10 & 82.32 \\
Qwen25-Coder-Instruct-33B & 84.15 & 78.66 & 76.22 & 76.83 \\
StarCoder2-15B & 71.34 & 64.02 & 60.98 & 60.98 \\
StarCoder2-7B & 31.10 & 25.61 & 23.17 & 24.39 \\
Qwen2.5-Coder-7B & 62.80 & 56.71 & 54.27 & 54.88 \\
\hline  
\end{tabular}
\end{table*}

To evaluate whether the effectiveness of the multi-task fine-tuning module depends on the specific backbone model, we instantiated the module with four alternative backbones: CodeLlama-13B, StarCoder2-15B, CodeLlama-7B, and Qwen2.5-Coder-7B. All other components of \textit{Multisage} were kept unchanged.

Table~\ref{tab:internal_validity_backbone} reports the translation success rates under these configurations.

Although absolute performance varies across backbone choices, the relative improvement trends remain consistent, indicating that \textit{Multisage} does not rely on a specific fine-tuning backbone.

\section{Supplementary Analysis on Java-to-Python Translation}
\label{appendix:full_results}
To provide a more comprehensive view of model-level performance, we present the detailed results for all evaluated models in Table~\ref{tab:full_results}. The table includes both baseline performance and results enhanced by \textit{Multisage}.

\begin{table}
\centering
\caption{Full results on Java-to-Python translation.}
\label{tab:full_results}
\renewcommand{\arraystretch}{1.2}
\setlength{\tabcolsep}{5pt}
\begin{tabular}{lcc}
\hline
\textbf{Model} & \textbf{Succ. (\%)} & \textbf{CodeBLEU (\%)} \\
\hline

\multicolumn{3}{l}{\textit{Specialized Translation Models}} \\
TransCoder & 28.05 & 61.24 \\
DOBF & 31.71 & 64.87 \\
TransCoder-ST & 40.24 & 60.09 \\
TransCoder-IR & 45.12 & 55.63 \\
INTERTRANS & 67.07 & 34.51 \\

\hline
\multicolumn{3}{l}{\textit{Baseline LLMs}} \\
DeepSeek-V3 & 90.24 & 50.27 \\
GPT-4 Turbo & 78.05 & 45.83 \\
Qwen-Plus & 81.10 & 47.15 \\
Qwen2.5-Coder-33B & 86.59 & 48.39 \\
StarCoder2-15B & 58.54 & 36.42 \\
StarCoder2-7B & 23.17 & 21.76 \\
Qwen2.5-Coder-7B & 56.10 & 34.58 \\

\hline
\multicolumn{3}{l}{\textit{+ Multisage}} \\
DeepSeek-V3 & \textbf{96.34} & \textbf{57.34} \\
GPT-4 Turbo & \textbf{92.68} & \textbf{52.16} \\
Qwen-Plus & \textbf{90.85} & \textbf{55.48} \\
Qwen2.5-Coder-33B & \textbf{93.29} & \textbf{56.27} \\
StarCoder2-15B & \textbf{70.73} & \textbf{45.63} \\
StarCoder2-7B & \textbf{46.95} & \textbf{31.29} \\
Qwen2.5-Coder-7B & \textbf{69.51} & \textbf{43.10} \\

\hline
\end{tabular}
\end{table}

\end{document}